\documentclass[
    amsmath,
    amssymb,
    amsfonts,
    reprint,
    aps,
    floatfix,
    twocolumn,
    longbibliography,
    superscriptaddress
]{revtex4-1}

\usepackage[utf8]{inputenc}
\usepackage[T1]{fontenc}
\usepackage{physics}
\usepackage{graphicx}
\usepackage{bbm, bm}
\usepackage{float}
\usepackage{lipsum}
\usepackage{xurl}    % add this in the preamble

\usepackage{tabularx}  % in preamble
\usepackage{booktabs}   % nicer rules
\renewcommand{\selectlanguage}[1]{} % The language field in each bibitem is causing errors so let's just short-circuit it

\usepackage[breaklinks=true, colorlinks]{hyperref}
\hypersetup{allcolors=blue}
\usepackage{cleveref}

\usepackage[dvipsnames]{xcolor}
\usepackage{makecell}
\usepackage{upgreek}
\usepackage{nicefrac}

\newcommand{\rr}{{\vb{r}}}
\newcommand{\rR}{{\vb{R}}}

\DeclareMathOperator*{\EX}{\mathbb{E}}
\DeclareMathOperator*{\argmin}{argmin}

\begin{document}

\title{
    % Variational neural network ansatz for bosonic quantum field theories
    Neural network quantum states in the grand canonical ensemble
}

\author{Anton Hul}
\email{a.hul@uw.edu.pl}
\affiliation{Faculty of Physics, University of Warsaw, 02-093 Warsaw, Poland}

\author{Matija Medvidović}
\email{mmedvidovic@ethz.ch}
\affiliation{Institute for Theoretical Physics, ETH Zürich, 8093 Zürich, Switzerland}

\author{Juan Carrasquilla}
\affiliation{Institute for Theoretical Physics, ETH Zürich, 8093 Zürich, Switzerland}

\date{\today}

\begin{abstract}
    Variational Monte Carlo calculations have recently reached state-of-the-art accuracy in the approximation of ground state properties of quantum many-body systems. Making use of flexible neural quantum states and automatic differentiation has bypassed traditional computational obstacles such as reliance on basis sets. In this paper, we propose a neural quantum state architecture capable of representing symmetric bosonic wavefunctions in Fock space, enabling the study of systems with variable particle number. By supplementing our variational state with Monte Carlo sampling and geometric optimization, we demonstrate competitive variational energies across an array of one- and two-dimensional systems, converging to the physical boson number under a set chemical potential. Our approach enables accurate estimates of one-body reduced density matrices, opening access to observables such as condensate fractions and radial density profiles from first principles. Our method opens the door to numerical predictions of key measurable quantities in practical grand canonical systems.
\end{abstract}

\maketitle

\section{Introduction}
\label{sec:introduction}

A fundamental challenge in the classical simulation of quantum many-body systems stems from the computational complexity associated with the storage and manipulation of an exact representation of the quantum state, which scale exponentially with the number of particles in the system. This observation has motivated the development of a wide array of algorithms and efficient compression strategies for quantum many-body states with an eye on solving currently intractable problems in strongly correlated quantum matter~\cite{avellaStronglyCorrelatedSystems2013}. In recent years, neural quantum states (NQS)~\cite{carrasquillaHowUseNeural2021, carleoSolvingQuantumManybody2017, carleoMachineLearningPhysical2019, dawidMachineLearningQuantum2025} have emerged, offering a favorable tradeoff between expressive power and computational efficiency. Motivated by explosive progress within the field of artificial intelligence (AI), recent breakthroughs in variational Monte Carlo (VMC) have leveraged massively parallel modern hardware and algorithms to bypass traditional computational bottlenecks \cite{medvidovicNeuralnetworkQuantumStates2024, langeArchitecturesApplicationsReview2024}.

A key example of such a computational upgrade is performing calculations directly in first quantization. Removing the reliance on basis sets, modern variational calculations have reached unprecedented accuracy~\cite{wuVariationalBenchmarksQuantum2024} in difficult many-fermion calculations in quantum chemistry~\cite{pfauInitioSolutionManyelectron2020, hermannDeepneuralnetworkSolutionElectronic2020, glehnSelfAttentionAnsatzAbinitio2023, fosterInitioFoundationModel2025, hermannInitioQuantumChemistry2023}, electron gas systems~\cite{pesciaMessagepassingNeuralQuantum2024, kimNeuralnetworkQuantumStates2024, smithUnifiedVariationalApproach2024, louNeuralWaveFunctions2024}, nuclear physics~\cite{lovatoHiddennucleonsNeuralnetworkQuantum2022}, and even non-equilibrium dynamics~\cite{nysAbinitioVariationalWave2024, schmittSimulatingDynamicsCorrelated2025}. Rapid algorithmic advances continue to unlock new physical insights for quantum physics at scale.

Traditional approaches assume a fixed number of particles $N$. However, the chemical potential $\mu$ may induce phase transitions, control density fluctuations, and determine the thermodynamic stability for systems amenable to a description in terms of the grand canonical ensemble (GCE). Recent experiments with cold atoms~\cite{krukFluctuationsNumberAtoms2025}, photonic condensates in optical microcavities~\cite{schmittObservationGrandCanonicalNumber2014} or exciton-polariton condensates~\cite{byrnesExcitonPolaritonCondensates2014} display natural grand canonical statistics. The violation of particle number conservation in these systems makes them unsuitable for study using the canonical NQS-VMC pipeline.

Bosonic quantum field theories are a key research frontier for variational methods. First-quantized variational calculations on bosonic systems in Refs.~\cite{pesciaNeuralnetworkQuantumStates2022, bedaqueNeuralNetworkSolutions2024, freitasSynergyDeepNeural2023} report accurate results, testing predictions for prototypical bosonic models. A field-theoretic description has been proposed by Ref.~\cite{martynVariationalNeuralNetworkAnsatz2023} and has demonstrated accuracy in one-dimensional systems, with a full treatment of particle number fluctuations in the grand canonical formalism.

Discrete Bose-Hubbard Hamiltonians successfully describe deep optical lattices. However, they fundamentally fail to capture the physics of continuous 1D systems and shallow traps. Experimental observations of a one-dimensional Tonks-Girardeau gas \cite{kinoshitaObservationOneDimensionalTonksGirardeau2004} and the quantum Newton's cradle \cite{kinoshitaQuantumNewtonsCradle2006} have highlighted the rich correlated dynamics in the continuum. Modeling these systems from first principles necessitates flexible, continuous-space variational architectures capable of handling particle fluctuations in the grand canonical ensemble.

In this paper, we propose a form for the total variational wavefunction in Fock space. Exploiting the natural permutation equivariance of the transformer architecture~\cite{vaswaniAttentionAllYou2017, dosovitskiyImageWorth16x162020, glehnSelfAttentionAnsatzAbinitio2023, spragueVariationalMonteCarlo2024, viterittiTransformerVariationalWave2023}, we construct a symmetric and systematically improvable trial state. Due to the transformer masking mechanism, our state is capable of outputting amplitudes from many particle-number sectors using the same set of variational parameters. All our calculations are performed by sampling configurations directly in coordinate space, avoiding biases associated with single-particle basis sets.

Additionally, we propose a method to efficiently compute grand canonical one-body reduced density matrices, yielding precise numerical estimates of experimentally measurable observables, such as condensate fractions and single-particle densities. Crucially, our approach converges to an integer number of particles under a set chemical potential for particle-conserving Hamiltonians, revealing ensemble equivalence without unnecessary assumptions.

Our grand canonical variational quantum field state is tested against an array of bosonic Hamiltonians in one and two dimensions. The Lieb-Liniger~\cite{liebExactAnalysisInteracting1963} and Calogero-Sutherland~\cite{calogeroSolutionOneDimensionalNBody1971, sutherlandExactResultsQuantum1971, moserThreeIntegrableHamiltonian1975, khareQuantumManybodyProblem1997} models are used as accuracy benchmarks, where we show precise agreement with exact results. We treat harmonically trapped bosons with short-range interactions as well, extracting accurate variational wavefunctions and demonstrating the practical applicability of modern VMC to real-world systems.

\begin{figure}[t!]
    \centering
    \includegraphics[width=1.0\linewidth]{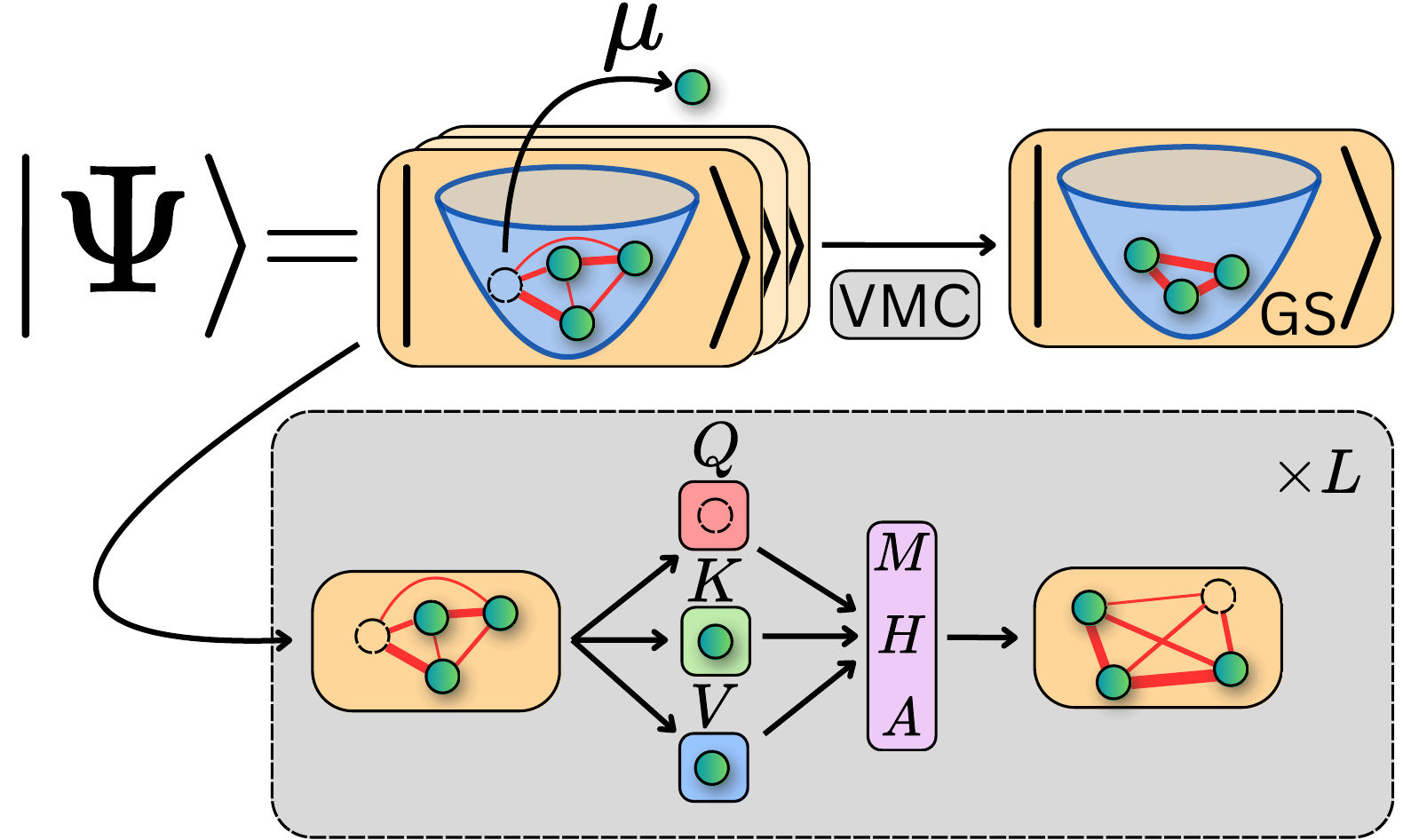}
    \caption{Overview of the TF based neural network ansatz, highlighting the masked multi-head attention (MHA) mechanism. The architecture is trained to approximate the ground state in the GCE using VMC optimization.}
    \label{fig:VMC_scheme}
\end{figure}

\section{Methods}
\label{sec:methods}
\subsection{Models}
Consider a general number-conserving Hamiltonian in the GCE, $H = H_1 + H_2$ with:
\begin{equation}
\begin{gathered}
    H_1 = \int _{\mathbb{R}^d} \dd{\rr} \, \hat \psi^\dag(\rr ) \left( -\frac{\nabla^2}{2m} + V(\rr )-\mu\right) \hat \psi(\rr ) \\
    H_2 = \frac{1}{2} \int _{\mathbb{R}^d} \dd{\rr} \dd{\rr'} \, W(\rr, \rr') \hat \psi^\dag(\rr )\hat \psi^\dag(\rr')\psi(\rr')\psi(\rr ) \; ,
    \label{eq:hamiltonian_grand_canonical}
\end{gathered}
\end{equation}
formulated as a $d$-dimensional spatial integral over one- and two-body terms, respectively. The Hamiltonian is expressed within the framework of second quantization, where the fields are represented by the creation and annihilation operators $\hat\psi^\dag(\rr)$ and $\hat\psi(\rr)$. It comprises the kinetic energy term, an external potential $V(\rr)$, and a two-body interaction potential $W(\rr, \rr')$. The particle number is not fixed but is instead determined by the chemical potential $\mu$.

In this framework, the quantum field states live in Fock space and can be expressed as a superposition of $n$-particle wave functions: 
\begin{equation}
    |\Psi\rangle = \bigoplus_{n=0}^{\infty} \int_{\mathbb{R}^{dn}} \dd{\rR_n}\, \phi_n(\rR_n)\,|\rR_n\rangle \, .
\end{equation}
As required in bosonic quantum field theories, the functions are permutation-invariant, unnormalized, and can be obtained by projecting the field state onto the corresponding basis state, $\phi_n(\rR_n) = \langle \rR_n|\Psi\rangle$.

\subsection{Variational ansatz}
\label{subsec:var_ansatz}

The variational ansatz is used to parametrize each $n$-particle wave function $\phi_n(\mathbf{R}_n)$ using a common neural network architecture. The architecture should be capable of handling an arbitrary or sufficiently large number of input arguments, i.e., particle positions, since the number of particles is not fixed.

The originally proposed ansatz for representing $\phi_n(\mathbf{R}_n)$ in one-dimensional systems relied on the Deep Sets (DS) neural network architecture~\cite{martynVariationalNeuralNetworkAnsatz2023}, which is permutation-invariant and naturally suited for inputs of variable size.

\begin{figure}[t!]
    \centering
    \includegraphics[width=\linewidth]{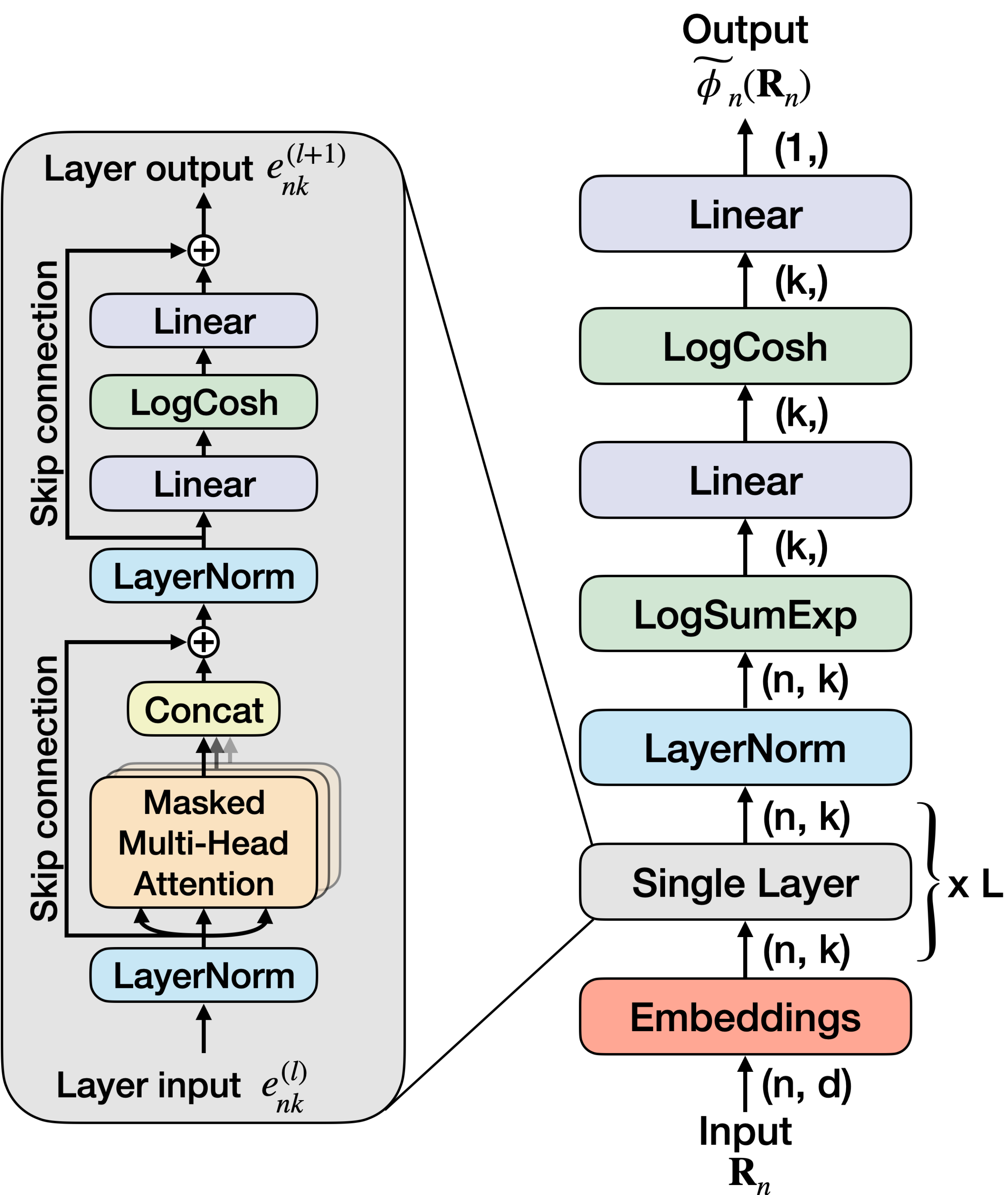}
    \caption{Schematic representation of the TF architecture ansatz. The particle positions $\vb{R}_n = \{\rr_i\}_{i=1}^n$ are input to evaluate the corresponding $n$-particle wavefunction $\varphi_n(\vb{R}_n)$, from which the full state $|\Psi\rangle$ can be calculated as a direct sum.}
    \label{fig:TF_architecture}
\end{figure}

In this work, we propose an alternative neural network ansatz based on the well-known transformer (TF) architecture~\cite{vaswaniAttentionAllYou2017}. As shown in Fig.~\ref{fig:TF_architecture}, the ansatz consists of $L$ identical transformer blocks. Each block includes a masked multi-head self-attention (MHA) layer with $H$ heads, preceded by layer normalization~\cite{baLayerNormalization2016}, and followed by a postprocessing sub-layer composed of layer normalization, linear layers, and activation functions. 

For a given input configuration $\rR_n \in \mathbb{R}^{n \times d}$, the MHA mechanism constructs three sets of matrices for each attention head $h$:
\begin{itemize}
    \item the queries $Q^{(h)} = \rR_n W_Q^{(h)} \in \mathbb{R}^{n \times d_k}$,
    \item the keys $K^{(h)} = \rR_n W_K^{(h)} \in \mathbb{R}^{n \times d_k}$,
    \item the values $V^{(h)} = \rR_n W_V^{(h)} \in \mathbb{R}^{n \times d_v}$,
\end{itemize}
where $W_Q^{(h)}$, $W_K^{(h)} \in \mathbb{R}^{d \times d_k}$, and $W_V^{(h)} \in \mathbb{R}^{d \times d_v}$ are trainable projection matrices.

Given a mask $M \in \mathbb{R}^{n \times n}$ encoding the absence of particles at each input position, the attention weights for head $h$ are computed as
\begin{equation}
    A^{(h)} = \mathrm{softmax}\!\left( \frac{Q^{(h)} {K^{(h)}}^\top + M}{\sqrt{d_k}} \right) \in \mathbb{R}^{n \times n},
\end{equation}
where the softmax is applied row-wise. The weighted combination of values, $A^{(h)} V^{(h)}$, from all $H$ heads is then concatenated and projected back to the original feature dimension. The weights of the attention matrices and linear layers are shared across all particles, but are optimized independently for each transformer block.

Given a permutation matrix $P \in \mathbb{R}^{n \times n}$ acting on the particle indices, the permuted input configuration is $P \rR_n$. The queries and keys transform as
\begin{equation}
    Q^{(h)} {K^{(h)}}^\top 
    \;\mapsto\; (P Q^{(h)})(P K^{(h)})^\top 
    = P \, Q^{(h)} {K^{(h)}}^\top P^\top.
\end{equation}
After applying the row-wise softmax, the attention matrix satisfies $A^{(h)} \;\mapsto\; P A^{(h)} P^\top$. Consequently, the value aggregation transforms as
\begin{equation}
    A^{(h)} V^{(h)} 
    \;\mapsto\; (P A^{(h)} P^\top)(P V^{(h)}) 
    = P \bigl(A^{(h)} V^{(h)}\bigr).
\end{equation}
By construction, the MHA mechanism guarantees permutation equivariance of the output with respect to the particle positions $\rR_n$. We then apply a \texttt{LogSumExp} pooling operator to obtain a permutation-invariant global representation:
\begin{equation}
    \text{LogSumExp}(x) = \log\left( \sum_{i=1}^k a_i e^{x_i} \right). 
\end{equation}
A detailed specification of the trainable parameters, including the number of attention heads, transformer layers, and typical embedding dimensions, is provided in Appendix~\ref{appendix:architecture}.

In systems with closed boundary conditions, the single-particle coordinates $\rr_i \in \mathbb{R}^d$ are embedded into a higher $k$-dimensional space using physically motivated Gaussian functions:
\begin{equation}
    \rr_i \;\longrightarrow\;
    \exp\!\left( -\frac{| \Delta \rr_i |^2}{2\sigma^2} \right)
    \in \mathbb{R}^{m^d},
\end{equation}
where $m$ denotes the number of uniformly spaced grid points  along each of the $d$ physical dimensions, and 
$\Delta \mathbf{r}_i = |\rr_i - \mathbf{s} |$ is the distance between the single-particle coordinate and the grid point $\mathbf{s}$. The parameter $\sigma$, which determines the width of the Gaussian functions, is kept constant and is chosen such that each Gaussian spans several grid points. The resulting $d$-dimensional grid representation is subsequently flattened into a one-dimensional array of length $k = m^d$.

In the presence of periodic boundary conditions (PBCs), the coordinates are embedded into a truncated $L$-periodic Fourier basis~\cite{pesciaNeuralnetworkQuantumStates2022}:
\begin{equation}
    \rr_i \;\longrightarrow\;
    \left[
    \sin\!\left(\frac{2\pi j}{L} \rr_i\right),\;
    \cos\!\left(\frac{2\pi j}{L} \rr_i\right)
    \right]_{j=1}^{m}
    \in \mathbb{R}^{2md}.
\end{equation}
where $\sin(\rr_i)$ and $\cos(\rr_i)$ denote the component-wise application of the trigonometric functions to the vector $\rr_i$. The resulting representation is flattened into a one-dimensional array of length $k = 2md$.

The final $n$-particle wave function $\varphi_n(\rR_n)$ is obtained by multiplying the ansatz output $\tilde{\varphi}_n(\rR_n)$ by additional factors introduced to enhance convergence during optimization. These may include a cutoff factor, which enforces the vanishing of the wave function at the system boundaries in closed systems, or a Jastrow factor, which ensures the correct short-range behavior of the wave function (see Appendix~\ref{appendix:addit_factors} for further details).

\subsection{Variational Monte Carlo}
The standard VMC approach enables the estimation of expectation values of a generic quantum mechanical operator $Q$ of the form
\begin{equation}
    \langle Q \rangle = \frac{\langle \psi_\theta  |Q|\psi_\theta  \rangle}{\langle \psi_\theta |\psi_\theta \rangle}
    \label{eq:vmc_expectation}
\end{equation}
where $|\psi_\theta \rangle$ is a variational many-body state defined by amplitudes $\psi_\theta(\mathbf{x})$ in a basis $\mathbf{x}$. The direct calculation of expectation values in the numerator and denominator is generally intractable. However, they can be estimated efficiently via Monte Carlo integration.  

An unbiased estimator for the expectation value in Eq.~\eqref{eq:vmc_expectation} can be obtained as the empirical mean of $Q^{\text{loc}}_n(\mathbf{x})$ over $N_s$ configurations $\mathbf{x}_i$ sampled from $|\bar{\varphi}(\mathbf{x})|^2$:
\begin{equation}
    \langle Q \rangle = \mathbb{E}_{\mathbf{x} \sim |\bar{\varphi}(\mathbf{x})|^2} \bigl[ Q^{\text{loc}}_n(\mathbf{x}) \bigr] \approx \frac{1}{N_s} \sum_{i=1}^{N_s} Q^{\text{loc}}_n(\mathbf{x}_i),
\end{equation}
where the local observable $Q^{\text{loc}}_n(\mathbf{x})$ is defined as
\begin{equation}
    Q^{\text{loc}}_n(\mathbf{x}) = \frac{\langle \mathbf{x} | Q | \psi_\theta \rangle}{\langle \mathbf{x} | \psi_\theta \rangle},
\end{equation}
and the normalized probability density is
\begin{equation}
    |\bar{\varphi}(\mathbf{x})|^2 = \frac{|\psi_\theta(\mathbf{x})|^2}{\int d\mathbf{x}'\, |\psi_\theta(\mathbf{x}')|^2}.
\end{equation}

An analogous method for estimating observables in the context of QFTs was developed in Ref.~\cite{martynVariationalNeuralNetworkAnsatz2023}. The approach, referred to as VMC in Fock space, expresses the expectation value of an $n$-particle local observable function $Q^{\text{loc}}_n(\rR_n)$ for a quantum field state $| \Psi \rangle$, as
\begin{equation}
    \langle Q \rangle = \frac{\langle \Psi |Q|\Psi \rangle}{\langle \Psi|\Psi\rangle} =  \EX_{n \sim P_n} \EX_{\rR_n \sim |\bar \varphi_n|^2} \biggr[Q^{\text{loc}}_n(\rR_n) \biggr]\, ,
    \label{eq:exp_value}
\end{equation}
where the n-particle local observable of the configuration $\rR_n$ is
\begin{equation}
    Q^{\text{loc}}_n(\rR_n) = \frac{ \langle \rR_n | Q |  \Psi \rangle}{ \langle \rR_n | \Psi\rangle} 
\end{equation}
with the probability distribution of the particle number
\begin{equation}
    P_n = \frac{ \langle \varphi_n | \varphi_n\rangle}{ \langle \Psi|\Psi \rangle} = \frac{\int d^n \rr |\varphi_n(\rR_n)|^2}{\sum_{m=0}^\infty \int d^n \vb{r}' |\varphi_m(\rR'_m)|^2} 
    \label{eq:def_pn}
\end{equation}
and the probability distribution of the n-particle wave function
\begin{equation}
    |\bar \varphi_n(\rR_n)|^2 = \frac{|\varphi_n(\rR_n)|^2}{\int d^n \rr' |\varphi_n(\rR'_n)|^2}\, .
    \label{eq:def_phi_bar}
\end{equation}
The approach can be seen as a natural generalization of the traditional VMC, where the expectation value is extended to include both the distribution of the $n$-particle wave function and the particle number distribution.

\subsection{Markov Chain Monte Carlo sampling}
\label{sec:mcmc}

To obtain the samples $\{\rR_i: i = 1, \dots, N_s\}$ necessary for estimating observables in Eq.~\eqref{eq:exp_value}, it is common practice to employ well-established Markov Chain Monte Carlo (MCMC) samplers. In this work, we largely follow the sampling procedure outlined in Ref.~\cite{martynVariationalNeuralNetworkAnsatz2023}. 

As in standard MCMC algorithms, the procedure in Fock space begins by proposing a new configuration $\rR_n'$ from the current configuration $\rR_n$, which is then accepted or rejected according to the Metropolis–Hastings rule. In our scheme, new bosonic configurations $\rR_n'$ are proposed while also allowing the particle number $n$ to change. With probability $2p_\pm$, the particle number is either increased or decreased, each occurring with probability $p_\pm$. With probability $p_0 = 1 - 2p_\pm$, the proposed configuration preserves the particle number, and a uniform random displacement is applied to all particle positions:
\begin{equation}
    \rR_n \rightarrow \rR_n' = \rR_n + \bm{\zeta},
\end{equation}
where $\bm{\zeta} \sim U\left([ -\tfrac{w}{2}, \tfrac{w}{2} ]^d\right)$ is a $d$-dimensional uniform random vector of width $w$. 

The probabilities of individual configurations $\rR_n$, as well as the transition probabilities $\rR_n \rightarrow \rR_n'$, are given in detail in Appendix~\ref{appendix:mcmc_fock}. In practice, we run multiple MCMC chains in parallel, and estimate expectation values as empirical means over all collected samples $\rR_n$ with different particle numbers $n$. The uncertainty of the estimated expectation value is computed as the standard deviation of the corresponding means across the chains.

\subsection{Stochastic optimization}
The goal of VMC is to minimize the energy expectation value with respect to the trial state $|\psi_\theta\rangle$ by finding the optimal variational parameters 
\begin{equation}
    \theta_{\text{opt}} = \argmin_{\theta} E(\theta) = \argmin_{\theta} \frac{\langle \psi_\theta|H|\psi_\theta \rangle}{\langle \psi_\theta|\psi_\theta \rangle} \geq E_0 \, ,
\end{equation}
where $E_0$ is the true ground state energy. Standard gradient-based optimization methods can be employed to determine the optimal parameters $ \theta_{\text{opt}} $ by iteratively updating the parameters using the energy gradient with respect to the variational parameters $ g = \nabla_\theta E(\theta) $, starting from an initial guess. 

A more general optimization strategy, motivated by physical principles and based on imaginary-time evolution, is known as natural gradient descent (NG) or stochastic reconfiguration (SR)~\cite{sorellaGreenFunctionMonte1998}. Within this framework, the updated ansatz parameters $\theta'$ are obtained as
\begin{equation}
    \theta' = \theta - \eta S^{-1} g,
    \label{eq:sr_update}
\end{equation}
where $\eta$ denotes the learning rate, and $S$ is the quantum geometric tensor (QGT), also referred to as the quantum Fisher information matrix. The gradient and the QGT are given by~\cite{medvidovicNeuralnetworkQuantumStates2024}:
\begin{equation}
\label{eq:s_matrix_and_gradient}
\begin{gathered}
    S_{\mu \nu} = 2 \Re \left\{ \langle O_\mu^\dagger ( O_\nu - \langle O_\nu \rangle ) \rangle \right\} \; , \\
    g_{\mu} = 2 \Re \left\{ \langle O_\mu^\dagger ( E - \langle E \rangle ) \rangle \right\},
\end{gathered}
\end{equation}
where the indices $ \mu $ and $ \nu $ run over the $ N_p $ variational parameters. The expectation values $ \langle \cdot \rangle \equiv \langle \psi_\theta | \cdot | \psi_\theta \rangle / \langle \psi_\theta | \psi_\theta \rangle $ are taken with respect to the trial state $ |\psi_\theta\rangle $. The diagonal operators $O_\mu$ are defined as
\begin{equation}
    O_\mu |\psi_\theta \rangle = \partial_{\theta^\mu} |\psi_\theta \rangle \, .
\end{equation}

The gradient and QGT in Eq.~\eqref{eq:s_matrix_and_gradient} depend only on the local energy $ E^{\text{loc}}_n(\rR_n) $ and the local operator $ O^{\text{loc}}_\nu(\rR_n) $, which are sampled using the MCMC method described in Sec.~\ref{sec:mcmc}. This procedure enables efficient estimation of the gradient and QGT required for parameter optimization, without the need for direct evaluation of $ \nabla_\theta E(\theta) $. Moreover, by applying the Woodbury identity, the effective matrix dimension for inversion can be reduced from $ N_p \times N_p $ to $ N_s \times N_s $~\cite{chenEmpoweringDeepNeural2024}. Instead of performing an explicit matrix inversion, the corresponding system of linear equations is solved using a standard dense linear algebra solver.

\section{Results}
\label{sec:results}
In this section, we first compare the performance of the introduced TF ansatz with the originally developed DS ansatz. In Secs.~\ref{subsec:ll_model} and \ref{subsec:cs_model_1d}, both architectures are applied to the exactly solvable one-dimensional Lieb–Liniger and Calogero–Sutherland models, respectively. We then focus exclusively on the TF ansatz in two-dimensional systems, studying the exactly solvable two-dimensional Calogero–Sutherland model in Sec.~\ref{subsec:cs_model_2d}. Finally, in Sec.~\ref{subsec:harm_confined}, we apply the TF ansatz to a system of harmonically confined bosons with Gaussian interactions. We show that our ansatz successfully reproduces the results obtained by exact diagonalization (ED) for a small number of bosons and, finally, present the results for an analytically unsolvable model in the grand-canonical ensemble.

\begin{figure}[!t]
    \centering
    \includegraphics[width=1.0\linewidth]{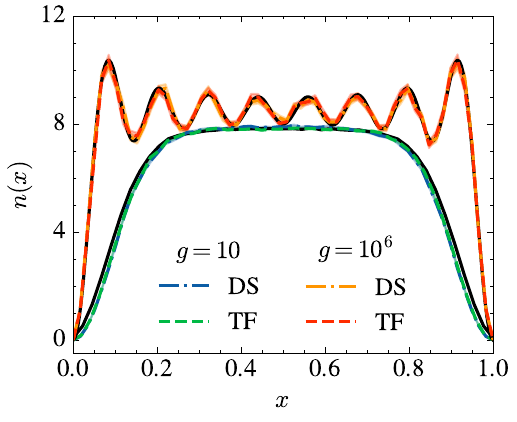}
    \caption{
        Particle number density $n(x)$ obtained from the converged ansatzes for the one-dimensional Lieb–Liniger model with $g = 10$ and $g = 10^6$. The exact particle number density is shown as a black line. The standard deviation is within the linewidth.
    }
    \label{fig:DS_TF_LL_1D}
\end{figure}

\subsection{Lieb-Liniger model}
\label{subsec:ll_model}
The original Lieb-Liniger model assumes the absence of an external potential $V(x)$ and describes an ensemble of $N$ bosons in a one-dimensional box of length $L$, interacting via a repulsive two-body contact potential:
\begin{equation}
    W_{\rm LL}(x, y) = 2g \, \delta(x-y),
\end{equation}  
with interaction strength $g > 0$. This model serves as a valuable numerical benchmark, allowing us to verify that both the DS and TF architectures can learn the ground state of a one-dimensional inhomogeneous system with short-range interactions and hard-wall boundaries, as well as to compare their performance. 

We set the particle mass $m = 1$ and the box length $L = 1$, and first consider the Tonks-Girardeau (TG) limit, in which the interaction strength is $g \to \infty$. We also examine the performance of both architectures at a smaller interaction strength of $g = 10$. The comparison of the ground-state energies and particle numbers obtained from the converged ansatzes is presented in Table~\ref{tab:compare_LL}, together with the corresponding exact values for two regimes. 

For a fair comparison, both models were configured with the same number of trainable parameters. While both architectures successfully converge to the correct ground-state energy, the TF architecture shows faster and smoother convergence and ultimately achieves a lower energy than the DS model. Moreover, the rescaled variance metric $\tilde{\sigma}^2_E$, which is known to correlate with the accuracy of the wave function (see Appendix~\ref{appendix:rescaled_var}), is one to two orders of magnitude smaller for the TF architecture.

To further verify the learned ground states, we calculate the particle number density, defined as:
\begin{equation}
    n(x) = \frac{\bra{\Psi} \hat \psi^\dag (x) \hat \psi (x) \ket{\Psi}}{\langle \Psi | \Psi \rangle}.
\end{equation}
As shown in Fig.~\ref{fig:DS_TF_LL_1D}, the densities estimated using VMC sampling (see Appendix~\ref{appendix:estim_obsv}) from the converged ansatzes perfectly match the exact analytical results.

\begin{figure}[!t]
    \centering
    \includegraphics[width=1.0\linewidth]{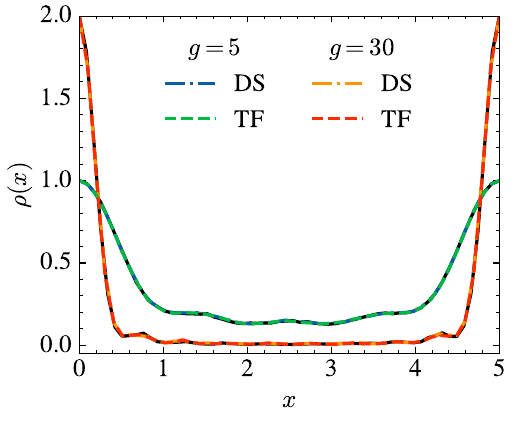}
    \caption{
    One-body density matrix $\rho(x)$ from the converged ansatzes applied to the one-dimensional Calogero–Sutherland model with $g = 5$ and $g = 30$. The exact one-body density matrix is shown as a black line. The standard deviation is within the linewidth.
    }
    \label{fig:DS_TF_CS_1D}
\end{figure}

\begin{table}[h]
\centering
\small
\caption{Comparison of ground-state energies and particle numbers for the one-dimensional Lieb-Liniger model obtained from the converged ansatzes, including the estimated rescaled variance metric and the corresponding exact values.}
\label{tab:compare_LL}
\setcellgapes{4pt} 
\makegapedcells
\begin{tabularx}{\columnwidth}{l *{2}{>{\centering\arraybackslash}X}}
\toprule
\textbf{Parameters} & \makecell[c]{$\mu = (8.75\pi)^2$ \\ $g = 10^6$} & \makecell[c]{$\mu = 115$ \\ $g = 10$} \\
\midrule
\textbf{Exact} & \makecell[c]{$E_0 = -4031.79$ \\ $n_0 = 8$} & \makecell[c]{$E_0 = -371.81$ \\ $n_0 = 6$} \\
\textbf{TF}    & \makecell[c]{$E = -4031.67 \pm 0.23$   \\ $n = 8.00 \pm 0.03$ \\ $\tilde{\sigma}^2_E = 5.01 \cdot 10^{-5}$} & \makecell[c]{$E = -371.78 \pm 0.08$  \\ $n = 6.00 \pm 0.08$ \\ $\tilde{\sigma}^2_E = 1.04 \cdot 10^{-3}$}  \\
\textbf{DS}    & \makecell[c]{$E = -4030.81 \pm 0.29$ \\ $n = 7.99 \pm 0.07 $ \\ $\tilde{\sigma}^2_E =  5.84\cdot 10^{-4} $} & \makecell[c]{$E = -371.23 \pm 0.07$ \\ $n = 6.00 \pm 0.13$ \\ $\tilde{\sigma}^2_E = 1.99 \cdot 10^{-2}$} \\
\bottomrule
\end{tabularx}
\end{table}

\subsection{Calogero-Sutherland model}
\label{subsec:cs_model}
\subsubsection{One-dimensional case}
\label{subsec:cs_model_1d}
Another valuable numerical benchmark for assessing whether both the DS and TF architectures can learn the ground state of a one-dimensional system with long-range interactions and periodic boundary conditions is the Calogero–Sutherland model. The model describes particles confined on a ring of circumference $L$, interacting via a two-body periodic potential that decays as the inverse square of the distance between particles:
\begin{equation}
    W_{\rm CS}(x,y) = g \frac{\pi^2}{L^2}\left[\sin \left(\frac{\pi}{L}(x-y)\right)\right]^{-2}
\end{equation}
Following the setup from Ref.~\cite{martynVariationalNeuralNetworkAnsatz2023}, we set the system size $L = 5$ and study both weaker and stronger coupling regimes. The comparison of the ground-state energies and particle numbers obtained from the converged ansatzes together with the corresponding exact values for two regimes is presented in Table~\ref{tab:compare_CS}. The TF architecture once again yields a rescaled variance metric $\tilde{\sigma}^2_E$ that is an order of magnitude smaller.

Another observable that can be calculated is the ground-state one-body density matrix (OBDM), defined as:
\begin{equation}
    \rho(x, x') = \frac{\bra{\Psi} \hat \psi^\dag (x') \hat \psi (x) \ket{\Psi}}{\langle \Psi | \Psi \rangle}.
\end{equation}
For the considered translation-invariant system, the OBDM depends only on the relative displacement, $\rho(s = |x - x'|)$. The resulting OBDMs estimated using VMC sampling (see Appendix~\ref{appendix:estim_obsv}) from the converged ansatzes are shown in Fig.~\ref{fig:DS_TF_CS_1D} and are in good agreement with the exact analytical results.

% \begin{figure}[!t]
%     \centering
%     \includegraphics[width=1.0\linewidth]{imgs/DS_TF_CS_1D.png}
%     \caption{
%     One-body density matrix $\rho(s)$ from the converged ansatzes applied to the one-dimensional Calogero–Sutherland model with $g = 5$ (left) and $g = 30$ (right). The exact one-body density matrix is shown as a black line.
%     }
%     \label{fig:DS_TF_CS_1D}
% \end{figure}

\begin{table}[h]
\centering
\small
\caption{Comparison of ground-state energies and particle numbers for the one-dimensional Calogero-Sutherland model obtained from the converged ansatzes, including the estimated rescaled variance metric and the corresponding exact values.}
\label{tab:compare_CS}
\setcellgapes{4pt} 
\makegapedcells
\begin{tabularx}{\columnwidth}{l *{2}{>{\centering\arraybackslash}X}}
\toprule
\textbf{Parameters} & \makecell[c]{$\mu = 3 \cdot 5^2 \cdot \frac{\pi^2 \lambda^2}{6 m L^2}$ \\ $g = 5$} & \makecell[c]{$\mu = 3 \cdot 10^2 \cdot \frac{\pi^2 \lambda^2}{6 m L^2}$ \\ $g = 30$} \\
\midrule
\textbf{Exact} & \makecell[c]{$E_0 = -156.317$ \\ $n_0 = 5$} & \makecell[c]{$E_0 = -5132.76 $ \\ $n_0 = 10$} \\
\textbf{TF}    & \makecell[c]{$E = -156.32 \pm 0.01$ \\ $n = 5 \pm 0.01$ \\ $\tilde{\sigma}^2_E =  2.03 \cdot 10^{-5}$} & \makecell[c]{$E = -5132.72 \pm 0.09$ \\ $n = 10.00 \pm 0.04$ \\ $\tilde{\sigma}^2_E = 1.36 \cdot 10^{-6}$} \\
\textbf{DS}    & \makecell[c]{$E = -156.32 \pm 0.01$ \\ $n = 5 \pm 0.02$ \\ $\tilde{\sigma}^2_E = 1.31 \cdot 10^{-4}$} & \makecell[c]{$E = -5132.53 \pm 0.03$ \\ $n = 9.99 \pm 0.12$ \\ $\tilde{\sigma}^2_E =  1.4 \cdot 10^{-5}$} \\
\bottomrule
\end{tabularx}
\end{table}

\subsubsection{Two-dimensional case}
\label{subsec:cs_model_2d}
\begin{figure}[!t]
    \centering
    \includegraphics[width=1.\linewidth]{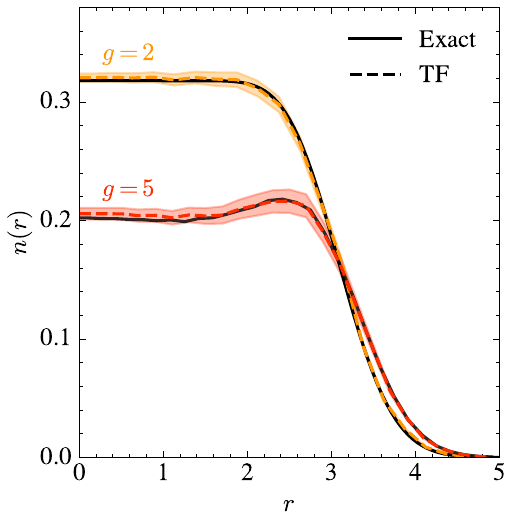}
    \caption{
    Particle number density profile $n(r)$ obtained from the converged ansatzes for the two-dimensional Calogero--Sutherland model with $g = 2.0$ and $g = 5.0$. The exact particle number density is shown as a black line. The standard deviation is indicated by the shaded region.}
    \label{fig:TF_CS_2D}
\end{figure}

An analytical extension of the original one-dimensional Calogero–Sutherland model to two dimensions is presented in Ref.~\cite{khareQuantumManybodyProblem1997}. The corresponding Hamiltonian describes a system of $N$ bosons confined within an isotropic harmonic potential,
\begin{equation}
    V(\rr) = \omega^2 \rr^2 ,
\end{equation}
and interacting through an inverse-square two-body potential,
\begin{equation}
    W^{(2)}_{\rm CS}(\rr, \rr') = \frac{g}{|\rr - \rr'|^2} ,
    \label{eq:harm_trap}
\end{equation}
as well as an additional three-body interaction,
\begin{equation}
    W^{(3)}_{\rm CS}(\rr, \rr', \rr'') = G \frac{\rr''- \rr}{|\rr''-\rr|^2} \cdot \frac{\rr''-\rr'}{|\rr''-\rr'|^2} ,
\end{equation}
where $\omega$ denotes the frequency of the harmonic trap, while $g$ and $G$ represent the coupling strengths of the two-body inverse-square and three-body interaction terms, respectively. The exact expression for the wave function density simplifies considerably for the specific choice of coupling constants, $g = G = 2$ (see Appendix~\ref{appendix:cs_2d}). Therefore, we focus on the model with this parameter setting, considering a system of size $L \times L$ with $L = 10$ and an angular frequency of $\omega = 1$. The resulting ground-state energies and particle numbers obtained from the converged TF ansatz, together with the corresponding exact values, are presented in Table~\ref{tab:compare_CS_2D}. The ground state particle number density profile $n(r \equiv |\rr|)$, obtained by averaging over the angular degrees of freedom, is shown in Fig.~\ref{fig:TF_CS_2D}, further confirming that the proposed TF architecture accurately captures the ground state of two-dimensional systems, even in the presence of three-body interactions.

\begin{figure}[t!]
    \centering
    \includegraphics[width=0.99\linewidth]{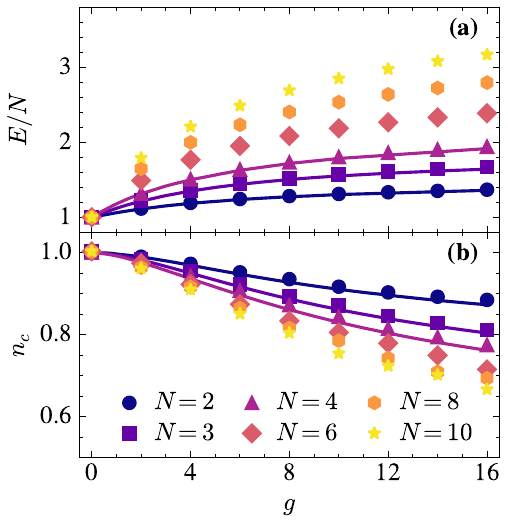}
    \caption{
        (a) Ground-state energy per particle $E/N$ and (b) condensate fraction $n_c$ for a system of $N$ two-dimensional harmonically confined bosons with Gaussian interactions of fixed range $s = 0.5$, across various interaction strengths $g$. Lines of different styles represent the ED results for a system of $N = 2, 3,$ and $4$, while markers correspond to the TF ansatz results. The standard deviation is smaller than the marker size.
    }
    \label{fig:canonical_gauss_E_dens}
\end{figure}

\begin{table}[h]
\centering
\small
\caption{Comparison of ground-state energies and particle numbers for the two-dimensional Calogero-Sutherland model obtained from the converged TF ansatz, including the estimated rescaled variance metric and the corresponding exact values.}
\label{tab:compare_CS_2D}
\setcellgapes{4pt} 
\makegapedcells
\begin{tabularx}{\columnwidth}{l *{3}{>{\centering\arraybackslash}X}}
\toprule
\textbf{Parameters} & \makecell[c]{$\mu = 22.0$ \\ g = G = 2.0} & \makecell[c]{$\mu = 25.0$ \\ g = G = 5.0} \\
\midrule
\textbf{Exact} & \makecell[c]{$E_0 = -110.0$ \\ $n_0 = 10$} & \makecell[c]{$E_0 = -95.46 $ \\ $n_0 = 8$} \\
\textbf{TF}    & \makecell[c]{$E = -109.98 \pm 0.01$ \\ $n = 10.00 \pm 0.06$ \\ $\tilde{\sigma}^2_E = 2.51 \cdot 10^{-6}$}  & \makecell[c]{$E = -95.44 \pm 0.01$ \\ $n = 7.99 \pm 0.09$ \\ $\tilde{\sigma}^2_E =   6.79 \cdot 10^{-6}$}  \\
\bottomrule
\end{tabularx}
\end{table}

\begin{figure*}[t!]
    \centering
\includegraphics[width=1.\linewidth]{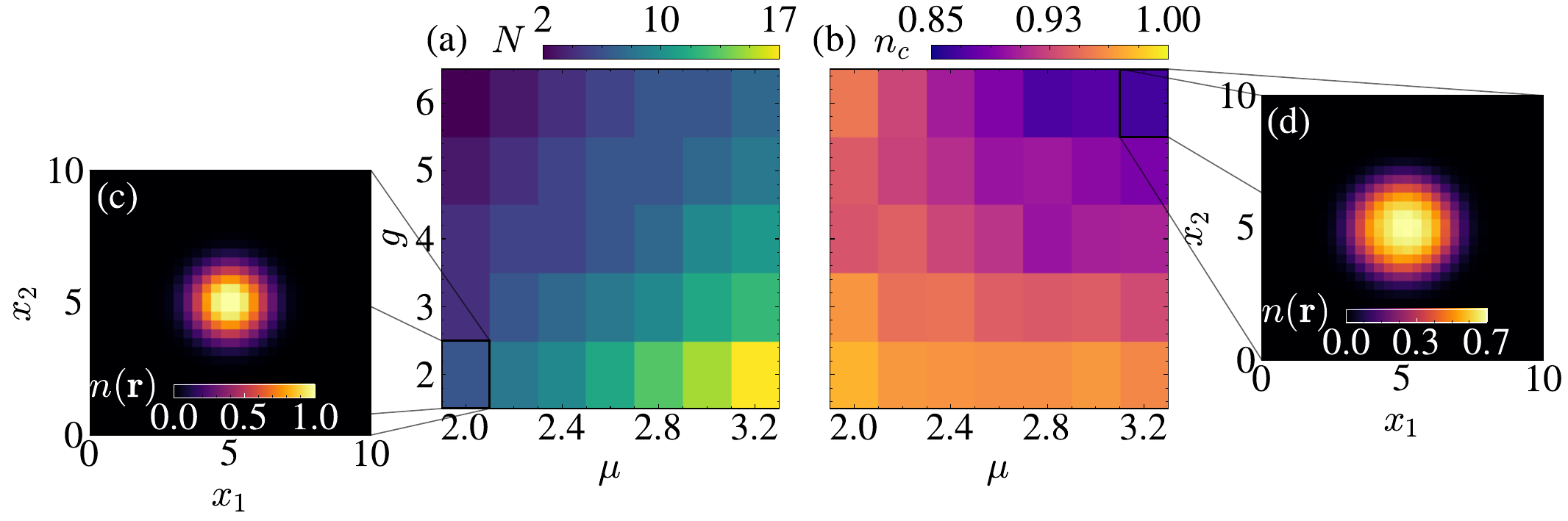}
    \caption{(a) Ground-state particle number $N$ and (b) condensate fraction $n_c$ for a system of two-dimensional harmonically confined bosons with Gaussian interactions of fixed range $s = 0.5$, shown as functions of the chemical potential $\mu$ and interaction strength $g$. Insets show the ground-state particle number density $n(\mathbf{r})$ for selected cases featuring (c) a small ($\mu = 2.0$, $g = 2.0$) and (d) a large ($\mu = 3.2$, $g = 6.0$) condensate fraction.}
\label{fig:grand_canonical_gauss_n_cond}
\end{figure*}

\subsection{Harmonically Confined Bosons with Gaussian Interactions}
\label{subsec:harm_confined}

We next consider a system of identical bosons confined in the two-dimensional isotropic harmonic trap defined in Eq.~\eqref{eq:harm_trap} and interacting via a Gaussian-shaped potential:
\begin{equation}
    W_{\rm GS}(\rr, \rr') = \frac{g}{\pi s^2} \, \mathrm{e}^{-\frac{|\rr - \rr'|^2}{s^2}},
\end{equation}
where $g$ denotes the interaction strength, and $s$ characterizes the range of the interaction.

Small systems with a fixed particle number, $N = 2$ to $N = 4$, can be exactly diagonalized as shown in Ref.~\cite{mujalQuantumCorrelationsDegeneracy2017}. By setting the probabilities $p_\pm$ for changing the particle number in the MCMC sampling algorithm (see Sec.~\ref{sec:mcmc}) to zero, the TF ansatz can be used to determine the ground state of the system within the canonical ensemble. As shown in Fig.~\ref{fig:canonical_gauss_E_dens}, the TF ansatz accurately reproduces the ground-state energies obtained from ED calculations.

The ground-state condensate fraction of the confined bosons can also be extracted as the largest eigenvalue of the OBDM~\cite{penroseBoseEinsteinCondensationLiquid1956a}. However, in the absence of translational invariance, the OBDM no longer depends solely on the relative displacement, making its estimation considerably more complex and computationally challenging (see Appendix~\ref{appendix:estim_obsv}). One possible approach to avoid its direct evaluation is to employ the so-called \textit{natural orbitals}, which play a role similar to marginal distributions for probability densities~\cite{pfauAccurateComputationQuantum2024}. The idea is to project the OBDM onto an orthogonal basis set of single-particle orbitals $\{\phi_1(\rr), \ldots, \phi_M(\rr)\}$, such that the elements of the projected OBDM, $\rho_{ij}$, are given by
\begin{equation}
    \rho_{ij} = \int \dd\rr \, \dd\rr' \, \phi_i(\rr)\,\phi_j(\rr')\,\rho(\rr, \rr') .
\end{equation}

A Monte Carlo estimator for $\rho_{ij}$ can then be written as~\cite{pfauAccurateComputationQuantum2024}:
\begin{equation}
    \rho_{ij} = 
    \EX_{\rr_n \sim |\bar \varphi_n|^2} 
    \EX_{\rr' \sim \rho} 
    \biggl[
    \frac{\phi_i(\rr)\,\phi_j(\rr')\,\varphi_n(\rr', \rr_2, \ldots, \rr_n)}
         {\rho(\rr')\,\varphi_n(\rr, \rr_2, \ldots, \rr_n)}
    \biggr] ,
    \label{eq:obdm_inhom_sampling}
\end{equation}
where $\rho$ is an arbitrary probability distribution over single-particle states. In practice, we take $\rho$ to be the ground-state density of a single harmonically trapped boson, while the natural orbitals $\ket{n_x,n_y}$ are chosen as the eigenstates of the two-dimensional harmonic oscillator.

The resulting condensate fraction obtained from this estimator is shown in Fig.~\ref{fig:canonical_gauss_E_dens}. The excellent agreement with the ED results demonstrates the accuracy and effectiveness of the OBDM projection approach. Crucially, the ansatz enables us to extend the analysis beyond ED limits, yielding results for systems of up to $N = 10$ bosons.

We next analyze the same system within the grand-canonical ensemble. By varying the chemical potential $\mu$ and the interaction strength $g$, while keeping the interaction range fixed at $s = 0.5$, we obtain the phase diagram shown in Fig.~\ref{fig:grand_canonical_gauss_n_cond}. As might be expected, the ground-state particle number increases with the chemical potential, and this growth is suppressed as the interaction strength is increased. In contrast, the condensate fraction remains relatively large over the entire range of $\mu$ for weak interactions but decreases significantly as $g$ and $\mu$ become large together. The ground state particle number density, displayed in the insets of Fig.~\ref{fig:grand_canonical_gauss_n_cond}, confirms that our ansatz accurately captures the ground state while preserving the required radial symmetry.

\section{Conclusion}
\label{sec:conclusion}

We have introduced a symmetric variational state for bosonic theories. The state makes use of the transformer architecture and its masking mechanism to parameterize the amplitudes of configurations with varying boson numbers using a universal set of variational parameters. Our trial wavefunction handles particle interactions using the attention mechanism in each block, offering a controllable and systematically improvable architecture. Because the state is particle-number agnostic, it can naturally be used for grand canonical \textit{open} systems, where equilibrium is maintained as particles are exchanged with a bath under a fixed chemical potential $\mu$. Focusing on many-boson systems in two dimensions, we demonstrate a computationally efficient VMC solver.

First-principles estimation of grand canonical states is a key ingredient in the study of systems where the phase diagram is probed by tuning the chemical potential, while the total number of particles may be difficult to control. Away from the thermodynamic limit, statistical ensembles are not equivalent, creating the need for an extension of the VMC toolbox to the grand canonical ensemble. The NQS pipeline is well-placed to address these \textit{ab-initio} computational challenges. Having become the default choice in most classical AI applications, variational states based on the transformer architecture offer a universal and rapidly scalable computational framework to reach accurate descriptions of these systems.

Equilibrium properties of such bosonic systems can be precisely evaluated Path Integral Monte Carlo and worm algorithms \cite{prokofevExactCompleteUniversal1998, boninsegniWormAlgorithmDiagrammatic2006, boninsegniWormAlgorithmContinuousSpace2006}. However, these traditional methods face severe limitations when computing real-time dynamics due to the dynamical sign problem \cite{cohenTamingDynamicalSign2015}. Explicit analytical trial states such as our transformer-based architecture can be extended using the Time-Dependent Variational Principle (TDVP). This clears the way to the study of real-time dynamics directly in continuous space.

Our variational approach is well suited to investigate cosmological phenomena simulated in ultracold atomic condensates, such as analog false vacuum decay \cite{bradenColdAtomAnalog2018}. Standard theoretical approaches to these setups rely on the Gross-Pitaevskii equation, which is limited by mean field truncation. Describing quantum tunneling and non-Gaussian fluctuations is difficult in the mean field language. On-demand access to accurate estimates of density profiles and the one-body density matrix can also be used for direct comparison with modern experiments \cite{bradenColdAtomAnalog2018, bottcherTransientSupersolidProperties2019, chomazLongLivedTransientSupersolid2019, chomazDipolarPhysicsReview2022}, helping to reveal new physics through parameter estimation and system control. Taking advantage of expressive modern trial states in VMC opens the door to the next generation of precision modeling of grand-canonical systems.

\section*{Software and simulations}

All simulations were performed on Nvidia L40 graphical processing units using the JAX~\cite{bradburyJAXComposableTransformations2018} library for array manipulation and automatic differentiation. Neural network design was conducted using Flax~\cite{heekFlaxNeuralNetwork2020}. Optax~\cite{deepmind2020jax} was used for optimization. NumPy~\cite{harrisArrayProgrammingNumPy2020} and SciPy~\cite{virtanenSciPy10Fundamental2020} were utilized for data processing, and plots were produced using Matplotlib~\cite{hunterMatplotlib2DGraphics2007}. The codes used to run the simulations and reproduce the examples presented in this paper are available on GitHub in Ref.~\cite{code}.

% \section{Data Availability}
% \label{sec:data}
% \mm{TO DO: replace with citation!}
% \mm{TO DO: should we write about the spiritbox resources?}

% \section*{Acknowledgements}
% \label{sec:acknowledgements}
% \mm{Do we have any?}

\bibliography{references-matija, references-tony}

@book{avellaStronglyCorrelatedSystems2013,
  title = {Strongly {{Correlated Systems}}: {{Numerical Methods}}},
  shorttitle = {Strongly {{Correlated Systems}}},
  editor = {Avella, Adolfo and Mancini, Ferdinando},
  year = 2013,
  series = {Springer {{Series}} in {{Solid-State Sciences}}},
  volume = {176},
  publisher = {Springer},
  address = {Berlin, Heidelberg},
  doi = {10.1007/978-3-642-35106-8},
  urldate = {2025-09-24},
  copyright = {https://www.springernature.com/gp/researchers/text-and-data-mining},
  isbn = {978-3-642-35105-1 978-3-642-35106-8},
  langid = {english}
}

@article{baLayerNormalization2016,
  title = {Layer {{Normalization}}},
  author = {Ba, Jimmy Lei and Kiros, Jamie Ryan and Hinton, Geoffrey E.},
  year = 2016,
  month = jul,
  eprint = {1607.06450},
  archiveprefix = {arXiv}
}

@book{beccaQuantumMonteCarlo2017,
  title = {Quantum {{Monte Carlo Approaches}} for {{Correlated Systems}}},
  author = {Becca, Federico and Sorella, Sandro},
  year = 2017,
  publisher = {Cambridge University Press},
  address = {Cambridge},
  doi = {10.1017/9781316417041},
  urldate = {2025-05-05},
  isbn = {978-1-107-12993-1}
}

@misc{bradburyJAXComposableTransformations2018,
  title = {{{JAX}}: Composable Transformations of {{Python}}+{{NumPy}} Programs},
  author = {Bradbury, James and Frostig, Roy and Hawkins, Peter and Johnson, Matthew James and Leary, Chris and Maclaurin, Dougal and Necula, George and Paszke, Adam and VanderPlas, Jake and {Wanderman-Milne}, Skye and Zhang, Qiao},
  year = 2018
}

@article{carleoMachineLearningPhysical2019,
  title = {Machine Learning and the Physical Sciences},
  author = {Carleo, Giuseppe and Cirac, Ignacio and Cranmer, Kyle and Daudet, Laurent and Schuld, Maria and Tishby, Naftali and {Vogt-Maranto}, Leslie and Zdeborov{\'a}, Lenka},
  year = 2019,
  month = dec,
  journal = {Reviews of Modern Physics},
  volume = {91},
  number = {4},
  eprint = {1903.10563},
  pages = {045002},
  publisher = {American Physical Society},
  issn = {15390756},
  doi = {10.1103/RevModPhys.91.045002},
  urldate = {2020-06-17},
  archiveprefix = {arXiv}
}

@article{carleoSolvingQuantumManybody2017,
  title = {Solving the Quantum Many-Body Problem with Artificial Neural Networks},
  author = {Carleo, Giuseppe and Troyer, Matthias},
  year = 2017,
  journal = {Science},
  volume = {355},
  number = {6325},
  pages = {602--606},
  issn = {10959203},
  doi = {10.1126/science.aag2302}
}

@article{carrasquillaHowUseNeural2021,
  title = {How {{To Use Neural Networks To Investigate Quantum Many-Body Physics}}},
  author = {Carrasquilla, Juan and Torlai, Giacomo},
  year = 2021,
  month = nov,
  journal = {PRX Quantum},
  volume = {2},
  number = {4},
  pages = {040201},
  publisher = {American Physical Society},
  doi = {10.1103/PRXQuantum.2.040201}
}

@article{chenEmpoweringDeepNeural2024,
  title = {Empowering Deep Neural Quantum States through Efficient Optimization},
  author = {Chen, Ao and Heyl, Markus},
  year = 2024,
  month = sep,
  journal = {Nature Physics},
  volume = {20},
  number = {9},
  pages = {1476--1481},
  publisher = {Nature Publishing Group},
  issn = {1745-2481},
  doi = {10.1038/s41567-024-02566-1},
  urldate = {2025-05-18},
  copyright = {2024 The Author(s)},
  langid = {english}
}

@book{dawidMachineLearningQuantum2025,
  title = {Machine Learning in Quantum Sciences},
  author = {Dawid, Anna and Arnold, Julian and Requena, Borja and Gresch, Alexander and P{\l}odzie{\'n}, Marcin and Donatella, Kaelan and Nicoli, Kim A. and Stornati, Paolo and Koch, Rouven and B{\"u}ttner, Miriam and Oku{\l}a, Robert and {Mu{\~n}oz-Gil}, Gorka and {Vargas-Hern{\'a}ndez}, Rodrigo A. and {Cervera-Lierta}, Alba and Carrasquilla, Juan and Dunjko, Vedran and Gabri{\'e}, Marylou and Huembeli, Patrick and van Nieuwenburg, Evert and Vicentini, Filippo and Wang, Lei and Wetzel, Sebastian J. and Carleo, Giuseppe and Greplov{\'a}, Eli{\v s}ka and Krems, Roman and Marquardt, Florian and Tomza, Micha{\l} and Lewenstein, Maciej and Dauphin, Alexandre},
  year = 2025,
  month = jun,
  eprint = {2204.04198},
  primaryclass = {quant-ph},
  publisher = {Cambridge University Press},
  urldate = {2025-09-30},
  archiveprefix = {arXiv}
}

@article{dosovitskiyImageWorth16x162020,
  title = {An {{Image}} Is {{Worth}} 16x16 {{Words}}: {{Transformers}} for {{Image Recognition}} at {{Scale}}},
  author = {Dosovitskiy, Alexey and Beyer, Lucas and Kolesnikov, Alexander and Weissenborn, Dirk and Zhai, Xiaohua and Unterthiner, Thomas and Dehghani, Mostafa and Minderer, Matthias and Heigold, Georg and Gelly, Sylvain and Uszkoreit, Jakob and Houlsby, Neil},
  year = 2020,
  month = oct,
  eprint = {2010.11929},
  archiveprefix = {arXiv}
}

@misc{fosterInitioFoundationModel2025,
  title = {An Ab Initio Foundation Model of Wavefunctions That Accurately Describes Chemical Bond Breaking},
  author = {Foster, Adam and Sch{\"a}tzle, Zeno and Szab{\'o}, P. Bern{\'a}t and Cheng, Lixue and K{\"o}hler, Jonas and Cassella, Gino and Gao, Nicholas and Li, Jiawei and No{\'e}, Frank and Hermann, Jan},
  year = 2025,
  month = jun,
  number = {arXiv:2506.19960},
  eprint = {2506.19960},
  primaryclass = {physics},
  publisher = {arXiv},
  doi = {10.48550/arXiv.2506.19960},
  urldate = {2025-08-19},
  archiveprefix = {arXiv}
}

@misc{glehnSelfAttentionAnsatzAbinitio2023,
  title = {A {{Self-Attention Ansatz}} for {{Ab-initio Quantum Chemistry}}},
  author = {von Glehn, Ingrid and Spencer, James S. and Pfau, David},
  year = 2023,
  month = apr,
  number = {arXiv:2211.13672},
  eprint = {2211.13672},
  primaryclass = {physics},
  publisher = {arXiv},
  doi = {10.48550/arXiv.2211.13672},
  urldate = {2025-09-30},
  archiveprefix = {arXiv}
}

@article{harrisArrayProgrammingNumPy2020,
  title = {Array Programming with {{NumPy}}},
  author = {Harris, Charles R. and Millman, K. Jarrod and {van der Walt}, St{\'e}fan J. and Gommers, Ralf and Virtanen, Pauli and Cournapeau, David and Wieser, Eric and Taylor, Julian and Berg, Sebastian and Smith, Nathaniel J. and Kern, Robert and Picus, Matti and Hoyer, Stephan and {van Kerkwijk}, Marten H. and Brett, Matthew and Haldane, Allan and {del R{\'i}o}, Jaime Fern{\'a}ndez and Wiebe, Mark and Peterson, Pearu and {G{\'e}rard-Marchant}, Pierre and Sheppard, Kevin and Reddy, Tyler and Weckesser, Warren and Abbasi, Hameer and Gohlke, Christoph and Oliphant, Travis E.},
  year = 2020,
  month = sep,
  journal = {Nature},
  volume = {585},
  number = {7825},
  eprint = {2006.10256},
  pages = {357--362},
  publisher = {Nature Research},
  issn = {14764687},
  doi = {10.1038/s41586-020-2649-2},
  urldate = {2020-11-03},
  archiveprefix = {arXiv},
  pmid = {32939066}
}

@misc{heekFlaxNeuralNetwork2020,
  title = {Flax: {{A}} Neural Network Library and Ecosystem for {{JAX}}},
  author = {Heek, Jonathan and Levskaya, Anselm and Oliver, Avital and Ritter, Marvin and Rondepierre, Bertrand and Steiner, Andreas and van Zee, Marc},
  year = 2020
}

@article{hermannDeepneuralnetworkSolutionElectronic2020,
  title = {Deep-Neural-Network Solution of the Electronic {{Schr\"odinger}} Equation},
  author = {Hermann, Jan and Sch{\"a}tzle, Zeno and No{\'e}, Frank},
  year = 2020,
  month = oct,
  journal = {Nature Chemistry},
  volume = {12},
  number = {10},
  pages = {891--897},
  issn = {1755-4330},
  doi = {10.1038/s41557-020-0544-y}
}

@article{hermannInitioQuantumChemistry2023,
  title = {Ab Initio Quantum Chemistry with Neural-Network Wavefunctions},
  author = {Hermann, Jan and Spencer, James and Choo, Kenny and Mezzacapo, Antonio and Foulkes, W. M. C. and Pfau, David and Carleo, Giuseppe and No{\'e}, Frank},
  year = 2023,
  month = oct,
  journal = {Nature Reviews Chemistry},
  volume = {7},
  number = {10},
  pages = {692--709},
  publisher = {Nature Publishing Group},
  issn = {2397-3358},
  doi = {10.1038/s41570-023-00516-8},
  urldate = {2026-01-27},
  copyright = {2023 Springer Nature Limited},
  langid = {english}
}

@article{hunterMatplotlib2DGraphics2007,
  title = {Matplotlib: {{A 2D}} Graphics Environment},
  author = {Hunter, John D.},
  year = 2007,
  journal = {Computing in Science and Engineering},
  volume = {9},
  number = {3},
  pages = {99--104},
  issn = {15219615},
  doi = {10.1109/MCSE.2007.55}
}

@article{kimNeuralnetworkQuantumStates2024,
  title = {Neural-Network Quantum States for Ultra-Cold {{Fermi}} Gases},
  author = {Kim, Jane and Pescia, Gabriel and Fore, Bryce and Nys, Jannes and Carleo, Giuseppe and Gandolfi, Stefano and {Hjorth-Jensen}, Morten and Lovato, Alessandro},
  year = 2024,
  month = may,
  journal = {Communications Physics},
  volume = {7},
  number = {1},
  pages = {148},
  publisher = {Nature Publishing Group},
  issn = {2399-3650},
  doi = {10.1038/s42005-024-01613-w},
  urldate = {2025-08-18},
  copyright = {2024 UChicago Argonne, LLC, Operator of Argonne National Laboratory},
  langid = {english}
}

@inproceedings{kingmaAdamMethodStochastic2015,
  title = {Adam: {{A}} Method for Stochastic Optimization},
  booktitle = {3rd {{International Conference}} on {{Learning Representations}}, {{ICLR}} 2015 - {{Conference Track Proceedings}}},
  author = {Kingma, Diederik P. and Ba, Jimmy Lei},
  year = 2015,
  month = dec,
  publisher = {International Conference on Learning Representations, ICLR}
}

@article{langeArchitecturesApplicationsReview2024,
  title = {From {{Architectures}} to {{Applications}}: {{A Review}} of {{Neural Quantum States}}},
  author = {Lange, Hannah and {Van de Walle}, Anka and Abedinnia, Atiye and Bohrdt, Annabelle},
  year = 2024,
  month = feb,
  eprint = {2402.09402},
  archiveprefix = {arXiv}
}

@article{lovatoHiddennucleonsNeuralnetworkQuantum2022,
  title = {Hidden-Nucleons Neural-Network Quantum States for the Nuclear Many-Body Problem},
  author = {Lovato, Alessandro and Adams, Corey and Carleo, Giuseppe and Rocco, Noemi},
  year = 2022,
  month = dec,
  journal = {Phys. Rev. Res.},
  volume = {4},
  number = {4},
  pages = {043178},
  publisher = {American Physical Society},
  doi = {10.1103/PhysRevResearch.4.043178}
}

@article{medvidovicNeuralnetworkQuantumStates2024,
  title = {Neural-Network Quantum States for Many-Body Physics},
  author = {Medvidovi{\'c}, Matija and Moreno, Javier Robledo},
  year = 2024,
  month = jul,
  journal = {The European Physical Journal Plus},
  volume = {139},
  number = {7},
  eprint = {2402.11014},
  primaryclass = {cond-mat},
  pages = {631},
  issn = {2190-5444},
  doi = {10.1140/epjp/s13360-024-05311-y},
  urldate = {2025-03-18},
  archiveprefix = {arXiv}
}

@article{nysAbinitioVariationalWave2024,
  title = {Ab-Initio Variational Wave Functions for the Time-Dependent Many-Electron {{Schr\"odinger}} Equation},
  author = {Nys, Jannes and Pescia, Gabriel and Sinibaldi, Alessandro and Carleo, Giuseppe},
  year = 2024,
  month = oct,
  journal = {Nature Communications},
  volume = {15},
  number = {1},
  eprint = {2403.07447},
  primaryclass = {cond-mat},
  pages = {9404},
  issn = {2041-1723},
  doi = {10.1038/s41467-024-53672-w},
  urldate = {2025-04-02},
  archiveprefix = {arXiv}
}

@article{pesciaMessagepassingNeuralQuantum2024,
  title = {Message-Passing Neural Quantum States for the Homogeneous Electron Gas},
  author = {Pescia, Gabriel and Nys, Jannes and Kim, Jane and Lovato, Alessandro and Carleo, Giuseppe},
  year = 2024,
  month = jul,
  journal = {Physical Review B},
  volume = {110},
  number = {3},
  pages = {035108},
  issn = {2469-9950, 2469-9969},
  doi = {10.1103/PhysRevB.110.035108},
  urldate = {2024-10-25},
  langid = {english}
}

@article{pesciaNeuralnetworkQuantumStates2022,
  title = {Neural-Network Quantum States for Periodic Systems in Continuous Space},
  author = {Pescia, Gabriel and Han, Jiequn and Lovato, Alessandro and Lu, Jianfeng and Carleo, Giuseppe},
  year = 2022,
  month = may,
  journal = {Physical Review Research},
  volume = {4},
  number = {2},
  pages = {023138},
  issn = {2643-1564},
  doi = {10.1103/PhysRevResearch.4.023138}
}

@article{pfauAccurateComputationQuantum2024,
  title = {Accurate Computation of Quantum Excited States with Neural Networks},
  author = {Pfau, David and Axelrod, Simon and Sutterud, Halvard and Von Glehn, Ingrid and Spencer, James S.},
  year = 2024,
  month = aug,
  journal = {Science},
  volume = {385},
  number = {6711},
  pages = {eadn0137},
  issn = {0036-8075, 1095-9203},
  doi = {10.1126/science.adn0137},
  urldate = {2025-08-19},
  langid = {english}
}

@article{pfauInitioSolutionManyelectron2020,
  title = {Ab Initio Solution of the Many-Electron {{Schr\"odinger}} Equation with Deep Neural Networks},
  author = {Pfau, David and Spencer, James S. and Matthews, Alexander G. D. G. and Foulkes, W. M. C.},
  year = 2020,
  month = sep,
  journal = {Physical Review Research},
  volume = {2},
  number = {3},
  pages = {033429},
  issn = {2643-1564},
  doi = {10.1103/PhysRevResearch.2.033429},
  urldate = {2025-09-30},
  langid = {english}
}

@misc{schmittSimulatingDynamicsCorrelated2025,
  title = {Simulating Dynamics of Correlated Matter with Neural Quantum States},
  author = {Schmitt, Markus and Heyl, Markus},
  year = 2025,
  month = jun,
  number = {arXiv:2506.03124},
  eprint = {2506.03124},
  primaryclass = {quant-ph},
  publisher = {arXiv},
  doi = {10.48550/arXiv.2506.03124},
  urldate = {2025-09-01},
  archiveprefix = {arXiv}
}

@article{smithUnifiedVariationalApproach2024,
  title = {Unified {{Variational Approach Description}} of {{Ground-State Phases}} of the {{Two-Dimensional Electron Gas}}},
  author = {Smith, Conor and Chen, Yixiao and Levy, Ryan and Yang, Yubo and Morales, Miguel A. and Zhang, Shiwei},
  year = 2024,
  month = dec,
  journal = {Physical Review Letters},
  volume = {133},
  number = {26},
  pages = {266504},
  publisher = {American Physical Society},
  doi = {10.1103/PhysRevLett.133.266504},
  urldate = {2025-08-18}
}

@article{sorellaGreenFunctionMonte1998,
  title = {Green Function Monte Carlo with Stochastic Reconfiguration},
  author = {Sorella, Sandro},
  year = 1998,
  month = may,
  journal = {Physical Review Letters},
  volume = {80},
  number = {20},
  eprint = {cond-mat/9803107},
  pages = {4558--4561},
  publisher = {American Physical Society},
  issn = {10797114},
  doi = {10.1103/PhysRevLett.80.4558},
  urldate = {2020-06-09},
  archiveprefix = {arXiv}
}

@article{vaswaniAttentionAllYou2017,
  title = {Attention {{Is All You Need}}},
  author = {Vaswani, Ashish and Shazeer, Noam and Parmar, Niki and Uszkoreit, Jakob and Jones, Llion and Gomez, Aidan N. and Kaiser, Lukasz and Polosukhin, Illia},
  year = 2017,
  month = jun,
  eprint = {1706.03762},
  archiveprefix = {arXiv}
}

@article{virtanenSciPy10Fundamental2020,
  title = {{{SciPy}} 1.0: Fundamental Algorithms for Scientific Computing in {{Python}}},
  shorttitle = {{{SciPy}} 1.0},
  author = {Virtanen, Pauli and Gommers, Ralf and Oliphant, Travis E. and Haberland, Matt and Reddy, Tyler and Cournapeau, David and Burovski, Evgeni and Peterson, Pearu and Weckesser, Warren and Bright, Jonathan and {van der Walt}, St{\'e}fan J. and Brett, Matthew and Wilson, Joshua and Millman, K. Jarrod and Mayorov, Nikolay and Nelson, Andrew R. J. and Jones, Eric and Kern, Robert and Larson, Eric and Carey, C. J. and Polat, {\.I}lhan and Feng, Yu and Moore, Eric W. and VanderPlas, Jake and Laxalde, Denis and Perktold, Josef and Cimrman, Robert and Henriksen, Ian and Quintero, E. A. and Harris, Charles R. and Archibald, Anne M. and Ribeiro, Ant{\^o}nio H. and Pedregosa, Fabian and {van Mulbregt}, Paul},
  year = 2020,
  month = mar,
  journal = {Nature Methods},
  volume = {17},
  number = {3},
  pages = {261--272},
  publisher = {Nature Publishing Group},
  issn = {1548-7105},
  doi = {10.1038/s41592-019-0686-2},
  urldate = {2025-09-25},
  copyright = {2020 The Author(s)},
  langid = {english}
}

@article{viterittiTransformerVariationalWave2023,
  title = {Transformer Variational Wave Functions for Frustrated Quantum Spin Systems},
  author = {Viteritti, Luciano Loris and Rende, Riccardo and Becca, Federico},
  year = 2023,
  month = jun,
  journal = {Physical Review Letters},
  volume = {130},
  number = {23},
  eprint = {2211.05504},
  primaryclass = {cond-mat},
  pages = {236401},
  issn = {0031-9007, 1079-7114},
  doi = {10.1103/PhysRevLett.130.236401},
  urldate = {2025-04-02},
  archiveprefix = {arXiv}
}

@article{wuVariationalBenchmarksQuantum2024,
  title = {Variational Benchmarks for Quantum Many-Body Problems},
  author = {Wu, Dian and Rossi, Riccardo and Vicentini, Filippo and Astrakhantsev, Nikita and Becca, Federico and Cao, Xiaodong and Carrasquilla, Juan and Ferrari, Francesco and Georges, Antoine and {Hibat-Allah}, Mohamed and Imada, Masatoshi and L{\"a}uchli, Andreas M. and Mazzola, Guglielmo and Mezzacapo, Antonio and Millis, Andrew and Robledo Moreno, Javier and Neupert, Titus and Nomura, Yusuke and Nys, Jannes and Parcollet, Olivier and Pohle, Rico and Romero, Imelda and Schmid, Michael and Silvester, J. Maxwell and Sorella, Sandro and Tocchio, Luca F. and Wang, Lei and White, Steven R. and Wietek, Alexander and Yang, Qi and Yang, Yiqi and Zhang, Shiwei and Carleo, Giuseppe},
  year = 2024,
  month = oct,
  journal = {Science},
  volume = {386},
  number = {6719},
  pages = {296--301},
  issn = {0036-8075, 1095-9203},
  doi = {10.1126/science.adg9774},
  urldate = {2025-03-07},
  langid = {english}
}

@article{martynVariationalNeuralNetworkAnsatz2023,
  title = {Variational {{Neural-Network Ansatz}} for {{Continuum Quantum Field Theory}}},
  author = {Martyn, John M. and Najafi, Khadijeh and Luo, Di},
  year = 2023,
  month = aug,
  journal = {Physical Review Letters},
  volume = {131},
  number = {8},
  pages = {081601},
  publisher = {American Physical Society},
  doi = {10.1103/PhysRevLett.131.081601},
  urldate = {2026-02-03}
}

@article{louNeuralWaveFunctions2024,
  title = {Neural {{Wave Functions}} for {{Superfluids}}},
  author = {Lou, Wan Tong and Sutterud, Halvard and Cassella, Gino and Foulkes, W. M. C. and Knolle, Johannes and Pfau, David and Spencer, James S.},
  year = 2024,
  month = may,
  journal = {Physical Review X},
  volume = {14},
  number = {2},
  pages = {021030},
  publisher = {American Physical Society},
  doi = {10.1103/PhysRevX.14.021030},
  urldate = {2026-02-03}
}

@article{byrnesExcitonPolaritonCondensates2014,
  title = {Exciton--Polariton Condensates},
  author = {Byrnes, Tim and Kim, Na Young and Yamamoto, Yoshihisa},
  year = 2014,
  month = nov,
  journal = {Nature Physics},
  volume = {10},
  number = {11},
  pages = {803--813},
  publisher = {Nature Publishing Group},
  issn = {1745-2481},
  doi = {10.1038/nphys3143},
  urldate = {2026-02-03},
  copyright = {2014 Springer Nature Limited},
  langid = {english}
}

@article{krukFluctuationsNumberAtoms2025,
  title = {On the Fluctuations of the Number of Atoms in the Condensate},
  author = {Kruk, Maciej B. and Kulik, Piotr and Andersen, Malthe F. and Deuar, Piotr and Gajda, Mariusz and Paw{\l}owski, Krzysztof and Witkowska, Emilia and Arlt, Jan J. and Rz{\k a}{\.z}ewski, Kazimierz},
  year = 2025,
  month = oct,
  journal = {Reports on Progress in Physics},
  volume = {88},
  number = {10},
  eprint = {2502.10880},
  primaryclass = {cond-mat},
  pages = {106401},
  issn = {0034-4885, 1361-6633},
  doi = {10.1088/1361-6633/ae0e33},
  urldate = {2026-02-03},
  archiveprefix = {arXiv}
}

@article{schmittObservationGrandCanonicalNumber2014,
  title = {Observation of {{Grand-Canonical Number Statistics}} in a {{Photon Bose-Einstein Condensate}}},
  author = {Schmitt, Julian and Damm, Tobias and Dung, David and Vewinger, Frank and Klaers, Jan and Weitz, Martin},
  year = 2014,
  month = jan,
  journal = {Physical Review Letters},
  volume = {112},
  number = {3},
  pages = {030401},
  publisher = {American Physical Society},
  doi = {10.1103/PhysRevLett.112.030401},
  urldate = {2026-02-03},
}

@article{bedaqueNeuralNetworkSolutions2024,
  title = {Neural Network Solutions of Bosonic Quantum Systems in One Dimension},
  author = {Bedaque, Paulo F. and Kumar, Hersh and Sheng, Andy},
  year = 2024,
  month = mar,
  journal = {Physical Review C},
  volume = {109},
  number = {3},
  pages = {034004},
  publisher = {American Physical Society},
  doi = {10.1103/PhysRevC.109.034004},
  urldate = {2026-02-03},
}

@article{freitasSynergyDeepNeural2023,
  title = {Synergy between Deep Neural Networks and the Variational Monte Carlo Method for Small ${}^4 \text{{He}} _{N} $ Clusters},
  author = {Freitas, William and Vitiello, S. A.},
  year = 2023,
  month = dec,
  journal = {Quantum},
  volume = {7},
  pages = {1209},
  publisher = {Verein zur F\"orderung des Open Access Publizierens in den Quantenwissenschaften},
  doi = {10.22331/q-2023-12-18-1209},
  urldate = {2026-02-03},
  langid = {british},
}

@article{calogeroSolutionOneDimensionalNBody1971,
  title = {Solution of the {{One}}-{{Dimensional N}}-{{Body Problems}} with {{Quadratic}} and/or {{Inversely Quadratic Pair Potentials}}},
  author = {Calogero, F.},
  year = 1971,
  month = mar,
  journal = {Journal of Mathematical Physics},
  volume = {12},
  number = {3},
  pages = {419--436},
  issn = {0022-2488},
  doi = {10.1063/1.1665604},
  urldate = {2026-02-04},
}

@article{khareQuantumManybodyProblem1997,
  title = {A Quantum Many-Body Problem in Two Dimensions: Ground State},
  shorttitle = {A Quantum Many-Body Problem in Two Dimensions},
  author = {Khare, Avinash and Ray, Koushik},
  year = 1997,
  month = jun,
  journal = {Physics Letters A},
  volume = {230},
  number = {3},
  pages = {139--143},
  issn = {0375-9601},
  doi = {10.1016/S0375-9601(97)00233-8},
  urldate = {2026-02-04}
}

@article{moserThreeIntegrableHamiltonian1975,
  title = {Three Integrable {{Hamiltonian}} Systems Connected with Isospectral Deformations},
  author = {Moser, J},
  year = 1975,
  month = may,
  journal = {Advances in Mathematics},
  volume = {16},
  number = {2},
  pages = {197--220},
  issn = {0001-8708},
  doi = {10.1016/0001-8708(75)90151-6},
  urldate = {2026-02-04}
}

@article{sutherlandExactResultsQuantum1971,
  title = {Exact {{Results}} for a {{Quantum Many-Body Problem}} in {{One Dimension}}},
  author = {Sutherland, Bill},
  year = 1971,
  month = nov,
  journal = {Physical Review A},
  volume = {4},
  number = {5},
  pages = {2019--2021},
  publisher = {American Physical Society},
  doi = {10.1103/PhysRevA.4.2019},
  urldate = {2026-02-04}
}

@article{liebExactAnalysisInteracting1963,
  title = {Exact {{Analysis}} of an {{Interacting Bose Gas}}. {{I}}. {{The General Solution}} and the {{Ground State}}},
  author = {Lieb, Elliott H. and Liniger, Werner},
  year = 1963,
  month = may,
  journal = {Physical Review},
  volume = {130},
  number = {4},
  pages = {1605--1616},
  publisher = {American Physical Society},
  doi = {10.1103/PhysRev.130.1605},
  urldate = {2026-02-04}
}

@software{code,
    author = {Hul, Anton},
    license = {Apache-2.0},
    title = {{Neural network quantum states in the grand canonical ensemble}},
    url = {https://github.com/AntonHul/Neural-network-quantum-states-in-the-grand-canonical-ensemble}
}

@software{deepmind2020jax,
  title = {The {D}eep{M}ind {JAX} {E}cosystem},
  author = {DeepMind and Babuschkin, Igor and Baumli, Kate and Bell, Alison and Bhupatiraju, Surya and Bruce, Jake and Buchlovsky, Peter and Budden, David and Cai, Trevor and Clark, Aidan and Danihelka, Ivo and Dedieu, Antoine and Fantacci, Claudio and Godwin, Jonathan and Jones, Chris and Hemsley, Ross and Hennigan, Tom and Hessel, Matteo and Hou, Shaobo and Kapturowski, Steven and Keck, Thomas and Kemaev, Iurii and King, Michael and Kunesch, Markus and Martens, Lena and Merzic, Hamza and Mikulik, Vladimir and Norman, Tamara and Papamakarios, George and Quan, John and Ring, Roman and Ruiz, Francisco and Sanchez, Alvaro and Sartran, Laurent and Schneider, Rosalia and Sezener, Eren and Spencer, Stephen and Srinivasan, Srivatsan and Stanojevi\'{c}, Milo\v{s} and Stokowiec, Wojciech and Wang, Luyu and Zhou, Guangyao and Viola, Fabio},
  url = {http://github.com/google-deepmind},
  year = {2020},
}

@article{spragueVariationalMonteCarlo2024,
  title = {Variational {{Monte Carlo}} with {{Large Patched Transformers}}},
  author = {Sprague, Kyle and Czischek, Stefanie},
  year = 2024,
  month = mar,
  journal = {Communications Physics},
  volume = {7},
  number = {1},
  eprint = {2306.03921},
  primaryclass = {quant-ph},
  pages = {90},
  issn = {2399-3650},
  doi = {10.1038/s42005-024-01584-y},
  urldate = {2026-02-10},
  archiveprefix = {arXiv},
}

@article{kinoshitaObservationOneDimensionalTonksGirardeau2004,
  title = {Observation of a {{One-Dimensional Tonks-Girardeau Gas}}},
  author = {Kinoshita, Toshiya and Wenger, Trevor and Weiss, David S.},
  year = 2004,
  month = aug,
  journal = {Science},
  volume = {305},
  number = {5687},
  pages = {1125--1128},
  publisher = {American Association for the Advancement of Science},
  doi = {10.1126/science.1100700},
  urldate = {2026-04-22}
}

@article{kinoshitaQuantumNewtonsCradle2006,
  title = {A Quantum {{Newton}}'s Cradle},
  author = {Kinoshita, Toshiya and Wenger, Trevor and Weiss, David S.},
  year = 2006,
  month = apr,
  journal = {Nature},
  volume = {440},
  number = {7086},
  pages = {900--903},
  issn = {0028-0836, 1476-4687},
  doi = {10.1038/nature04693},
  urldate = {2026-04-22},
  copyright = {http://www.springer.com/tdm},
  langid = {english}
}

@article{boninsegniWormAlgorithmContinuousSpace2006,
  title = {Worm {{Algorithm}} for {{Continuous-Space Path Integral Monte Carlo Simulations}}},
  author = {Boninsegni, Massimo and Prokof'ev, Nikolay and Svistunov, Boris},
  year = 2006,
  month = feb,
  journal = {Physical Review Letters},
  volume = {96},
  number = {7},
  pages = {070601},
  publisher = {American Physical Society},
  doi = {10.1103/PhysRevLett.96.070601},
  urldate = {2026-04-22},
}

@article{boninsegniWormAlgorithmDiagrammatic2006,
  title = {Worm Algorithm and Diagrammatic {{Monte Carlo}}: {{A}} New Approach to Continuous-Space Path Integral {{Monte Carlo}} Simulations},
  shorttitle = {Worm Algorithm and Diagrammatic {{Monte Carlo}}},
  author = {Boninsegni, M. and Prokof'ev, N. V. and Svistunov, B. V.},
  year = 2006,
  month = sep,
  journal = {Physical Review E},
  volume = {74},
  number = {3},
  pages = {036701},
  publisher = {American Physical Society},
  doi = {10.1103/PhysRevE.74.036701},
  urldate = {2026-04-22},
}

@article{cohenTamingDynamicalSign2015,
  title = {Taming the {{Dynamical Sign Problem}} in {{Real-Time Evolution}} of {{Quantum Many-Body Problems}}},
  author = {Cohen, Guy and Gull, Emanuel and Reichman, David R. and Millis, Andrew J.},
  year = 2015,
  month = dec,
  journal = {Physical Review Letters},
  volume = {115},
  number = {26},
  pages = {266802},
  publisher = {American Physical Society},
  doi = {10.1103/PhysRevLett.115.266802},
  urldate = {2026-04-22},
}

@article{prokofevExactCompleteUniversal1998,
  title = {Exact, Complete, and Universal Continuous-Time Worldline {{Monte Carlo}} Approach to the Statistics of Discrete Quantum Systems},
  author = {Prokof'ev, N. V. and Svistunov, B. V. and Tupitsyn, I. S.},
  year = 1998,
  month = aug,
  journal = {Journal of Experimental and Theoretical Physics},
  volume = {87},
  number = {2},
  pages = {310--321},
  issn = {1090-6509},
  doi = {10.1134/1.558661},
  urldate = {2026-04-22},
  langid = {english},
}

@article{bradenColdAtomAnalog2018,
  title = {Towards the Cold Atom Analog False Vacuum},
  author = {Braden, Jonathan and Johnson, Matthew C. and Peiris, Hiranya V. and Weinfurtner, Silke},
  year = 2018,
  month = jul,
  journal = {Journal of High Energy Physics},
  volume = {2018},
  number = {7},
  pages = {14},
  issn = {1029-8479},
  doi = {10.1007/JHEP07(2018)014},
  urldate = {2026-04-22},
  langid = {english}
}

@article{bottcherTransientSupersolidProperties2019,
  title = {Transient {{Supersolid Properties}} in an {{Array}} of {{Dipolar Quantum Droplets}}},
  author = {B{\"o}ttcher, Fabian and Schmidt, Jan-Niklas and Wenzel, Matthias and Hertkorn, Jens and Guo, Mingyang and Langen, Tim and Pfau, Tilman},
  year = 2019,
  month = mar,
  journal = {Physical Review X},
  volume = {9},
  number = {1},
  pages = {011051},
  publisher = {American Physical Society},
  doi = {10.1103/PhysRevX.9.011051},
  urldate = {2026-04-22}
}

@article{chomazDipolarPhysicsReview2022,
  title = {Dipolar Physics: A Review of Experiments with Magnetic Quantum Gases},
  shorttitle = {Dipolar Physics},
  author = {Chomaz, Lauriane and {Ferrier-Barbut}, Igor and Ferlaino, Francesca and {Laburthe-Tolra}, Bruno and Lev, Benjamin L and Pfau, Tilman},
  year = 2022,
  month = dec,
  journal = {Reports on Progress in Physics},
  volume = {86},
  number = {2},
  pages = {026401},
  publisher = {IOP Publishing},
  issn = {0034-4885},
  doi = {10.1088/1361-6633/aca814},
  urldate = {2026-04-22},
  langid = {english}
}

@article{chomazLongLivedTransientSupersolid2019,
  title = {Long-{{Lived}} and {{Transient Supersolid Behaviors}} in {{Dipolar Quantum Gases}}},
  author = {Chomaz, L. and Petter, D. and Ilzh{\"o}fer, P. and Natale, G. and Trautmann, A. and Politi, C. and Durastante, G. and {van Bijnen}, R. M. W. and Patscheider, A. and Sohmen, M. and Mark, M. J. and Ferlaino, F.},
  year = 2019,
  month = apr,
  journal = {Physical Review X},
  volume = {9},
  number = {2},
  pages = {021012},
  publisher = {American Physical Society},
  doi = {10.1103/PhysRevX.9.021012},
  urldate = {2026-04-22}
}

@article{mujalQuantumCorrelationsDegeneracy2017,
  title = {Quantum Correlations and Degeneracy of Identical Bosons in a Two-Dimensional Harmonic Trap},
  author = {Mujal, Pere and Sarl{\'e}, Enric and Polls, Artur and {Juli{\'a}-D{\'i}az}, Bruno},
  year = 2017,
  month = oct,
  journal = {Physical Review A},
  volume = {96},
  number = {4},
  pages = {043614},
  issn = {2469-9926, 2469-9934},
  doi = {10.1103/PhysRevA.96.043614},
  urldate = {2025-09-14},
  copyright = {https://link.aps.org/licenses/aps-default-license},
  langid = {english}
}

@article{penroseBoseEinsteinCondensationLiquid1956a,
  title = {Bose-{{Einstein Condensation}} and {{Liquid Helium}}},
  author = {Penrose, Oliver and Onsager, Lars},
  year = 1956,
  month = nov,
  journal = {Physical Review},
  volume = {104},
  number = {3},
  pages = {576--584},
  issn = {0031-899X},
  doi = {10.1103/PhysRev.104.576},
  urldate = {2025-09-14},
  copyright = {http://link.aps.org/licenses/aps-default-license},
  langid = {english}
}

@book{mehta1991random,
  title={Random Matrices},
  author={Mehta, Madan Lal},
  edition={2nd},
  year={1991},
  publisher={Academic Press},
  address={New York}
}

@article{foulkesQuantumMonteCarlo2001,
  title = {Quantum {{Monte Carlo}} Simulations of Solids},
  author = {Foulkes, W. M. C. and Mitas, L. and Needs, R. J. and Rajagopal, G.},
  year = 2001,
  month = jan,
  journal = {Reviews of Modern Physics},
  volume = {73},
  number = {1},
  pages = {33--83},
  publisher = {American Physical Society},
  doi = {10.1103/RevModPhys.73.33},
  urldate = {2026-02-20},
}

\onecolumngrid
\appendix

\section{Architecture}
\label{appendix:architecture}
In this work, we propose an alternative neural network ansatz, based on the well-known Transformer (TF) architecture, originally introduced as a deep learning model in 2017 \cite{vaswaniAttentionAllYou2017}. The core mechanism that learns correlations by assigning weights to each pair of inputs is the so-called \textit{self-attention} mechanism. This mechanism plays a particularly crucial role in our case, as it ensures the permutation invariance of the architecture with respect to its input -- a fundamental requirement for representing bosonic wavefunctions. A single layer of the TF model, as illustrated in Fig.~\ref{fig:TF_architecture}, contains a masked multi-head self-attention component composed of $H$ independent self-attention heads. Each head is parameterized independently and operates on the input in parallel, allowing the model to learn correlations across multiple scales, while a mask is used to exclude inputs corresponding to non-existent particles. The output of the multi-head self-attention component is obtained by concatenating the outputs of all individual attention heads, resulting in a tensor of the same shape $(n_{\rm max}, k)$ as the input. To enhance the representational power of the model, the multi-head self-attention block is followed by linear layers, which act independently on each input position. Additionally, to ensure the stability of the model, skip connections and layer normalization are employed in each layer.

The final TF architecture is illustrated in Fig.~\ref{fig:TF_architecture}. Initially, the input is transformed using an embedding function and is then processed by a stack of $L$ identical TF layers. Each layer consists of a MHA sub-layer with $H$ heads. The weights of the MHA matrices and the linear layers are shared between all particle positions but are independently learned for each layer $l$. The exact specification of trainable parameters, including the number of attention heads, transformer layers, and the typical embedding dimension, is summarized in Table~\ref{tab:tf_params}. In practice, we typically choose $L = 2$ and $H = 4$.
\begin{table}[H]
    \centering
    \begin{tabular}{|c|c|}
        \hline
        \textbf{Parameter} & \textbf{Value} \\ \hline
        Embedding length $k$ & 100 \\ \hline
        Number of transformer layers $L$ & 2 \\ \hline
        Number of self-attention heads $H$ & 4 \\ \hline
        Width of the feedforward network & 100 \\ \hline
        Total number of trainable parameters & $\sim 130{,}000$ \\ \hline
    \end{tabular}
    \caption{Summary of TF architecture parameters.}
    \label{tab:tf_params}
\end{table}

After passing through all TF layers, the output tensor retains the same shape as the input to the first TF layer, that is, $(n, k)$. To obtain the $n$-particle wavefunction $\varphi_n(\rR_n)$, this tensor must be projected into a single scalar value. We begin by reducing along the first dimension, which corresponds to the number of particles, using the \textit{LogSumExp} function. The resulting vector of shape $(k,)$ is then mapped to a scalar through a feed-forward neural network.
\begin{equation}
\text{LogSumExp}(x) = \log\left( \sum_{i=1}^k a_i e^{x_i} \right)
\label{eq:logsumexp}
\end{equation}
It is worth noting that throughout the entire TF architecture, we employ the \textit{LogCosh} activation function, in contrast to the \textit{Tanh} activation function used in the DS architecture. The network weights are initialized using the Xavier uniform initialization method, which is designed to maintain the variance of activations across layers. Bias terms are initialized to zero throughout the network.

\section{Additional factors}
\label{appendix:addit_factors}
The additional factors must be incorporated into the ansatz to facilitate its optimization. In the following section, we analyze three categories of such factors, originally introduced in Ref.~\cite{martynVariationalNeuralNetworkAnsatz2023}. Some of these enable the incorporation of physical constraints of the system under study, while others are intended to enhance convergence during the optimization.

\subsection{Cutoff Factor}
The cutoff factor is used to enforce closed boundary conditions within the system. These conditions require that the $n$-particle wave function vanishes as any particle approaches the boundary of the system. To satisfy this constraint, the ansatz is multiplied by a factor that ensures the wave function tends to zero at the boundaries. 

In the one-dimensional case, where the system is defined on the interval $(0, L)$, the cutoff factor takes the following form:
\begin{equation}
    \left( \frac{L}{30} \right)^{-n/2} \prod_{i=1}^{n} \frac{x_i}{L} \left( 1 - \frac{x_i}{L} \right).
\end{equation}
In the two-dimensional case, the system is defined on the square domain $(0, L) \times (0, L)$, and the cutoff factor is accordingly adapted to vanish at all boundaries:
\begin{equation}
    \left( \frac{L}{100} \right)^{-n} \prod_{i=1}^{n} \frac{x_i}{L} \left( 1 - \frac{x_i}{L} \right)\prod_{i=1}^{n} \frac{y_i}{L} \left( 1 - \frac{y_i}{L} \right),
\end{equation}
where single particle position $\rr_i = (x_i, y_i)$ and the normalization in front is included for stable optimization across different particle number sectors~\cite{martynVariationalNeuralNetworkAnsatz2023}. In the case of periodic boundary conditions, the periodicity of the system is already incorporated into the embedding function (see Sec.~\ref{subsec:var_ansatz}). Therefore, no additional cutoff factor is required.

\subsection{Particle number distribution factor}
We additionally incorporate a parameterizable regularization factor $q_n$ into the $n$-particle wave function to ensure a well-behaved particle-number distribution. This factor is defined as a smoothed rectangular pulse that decays exponentially outside a central window:
\begin{equation}
q_n = \frac{1}{1 + e^{-s(n - c_1)}} \cdot \frac{1}{1 + e^{s(n - c_2)}}, \quad 0 \leq c_1 \leq c_2, \quad s > 0.
\end{equation}
\noindent
The purpose of this factor is to confine the most probable values of the particle number $n$ to the finite interval $[c_1, c_2]$. In practice, the parameters $ c_1 $, $ c_2 $, and $ s $ are optimized together with the rest of the ansatz parameters using a gradient-based optimization algorithm, typically with a higher learning rate.

\subsection{Jastrow Factor}

In some of the models considered in this work, the two-body interaction potential $W(\rr_i, \rr_j)$ diverges for a configuration $\rR_n$ in which two particles approach one another, $\rr_i \rightarrow \rr_j$. For the exact ground-state wave function, this divergence is typically canceled by a corresponding divergence in the kinetic energy. However, since the ansatz is only an approximation to the true ground state, the local energy $E^{\text{loc}}_n(\rR_n)$ for such configurations may still exhibit divergences. These divergences can lead to large oscillations in the estimated energy during MCMC sampling, thereby complicating convergence to the true ground state.

To address this issue, we incorporate the correct short-range cusp behavior into the variational ansatz, thereby enforcing Kato’s cusp condition~\cite{foulkesQuantumMonteCarlo2001}, which ensures that the local energy remains finite as $\rr_i \rightarrow \rr_j$. This can be achieved without explicit knowledge of the exact many-body ground state by analyzing the corresponding two-body problem and imposing the requirement:
\begin{equation}
    \lim_{\rr_i \rightarrow \rr_j} E^{\text{loc}}_n(\rR_n)  =  \lim_{\rr_i \rightarrow \rr_j} \frac{H\varphi_n(\rR_n)}{\varphi_n(\rR_n)} < \infty.
\end{equation}
The Jastrow factors for the Lieb–Liniger model and the one-dimensional Calogero–Sutherland model were derived in Ref.~\cite{martynVariationalNeuralNetworkAnsatz2023}. For the reader’s convenience, we present only the final expressions here:
\begin{equation}
    J^{\text{LL}}_n(X_n) = \prod_{i < j}^{n} \left( \frac{1}{L} |x_i - x_j| + \frac{1}{mgL} \right),
\end{equation}
\begin{equation}
    J^{\text{CS}}_n(X_n) = \prod_{i < j}^{n} \left( \frac{1}{L} |x_i - x_j|\right)^\lambda \cdot \left(1- \frac{1}{L} |x_i - x_j|\right)^\lambda,
\end{equation}
where $ \lambda = \frac{1}{2}\biggr(1 + \sqrt{1 + 4mg}\biggr)$.

The Jastrow factor for the inverse-square two-body potential $
W(\rr, \rr') = g/|\rr - \rr'|^2 $ in the two-dimensional Calogero–Sutherland model can be derived in an analogous manner. We begin by considering only the radial part of the Laplacian expressed in polar coordinates:
\begin{equation}
\nabla^2 f = \frac{\partial^2 f}{\partial r^2} + \frac{1}{r} \frac{\partial f}{\partial r}.
\end{equation}
The corresponding cusp condition for the inverse-square interaction then takes the form
\begin{equation}
-\left(\frac{\partial^2}{\partial r^2} + \frac{1}{r} \frac{\partial}{\partial r}\right) u(|\rr|) + g \frac{u(|\rr|)}{r^2} = 0.
\end{equation}
This condition is satisfied by the trial function $u(|\rr|) = |\rr|^\lambda$ with the choice $\lambda = \sqrt{mg} = \sqrt{g/2}$. Consequently, the Jastrow factor for the two-dimensional Calogero–Sutherland model is given by
\begin{equation}
J^{\text{CS}}_n(\rR_n) = \prod_{i < j}^n |\rr_i - \rr_j|^\lambda.
\end{equation}
Although our derivation of the Jastrow factor accounts only for two-body inverse-square interactions and neglects the stronger three-body divergence that arises when three particles simultaneously approach one another, the resulting factor is nevertheless sufficient in practice to suppress large oscillations in the local energy.

\section{VMC in Fock space}
\label{appendix:vmc_fock_space}
In a standard Variational Monte Carlo (VMC), the characterization of the physical properties of many-body system requires the efficient computation of expectation values of the form:
\begin{equation}
    \langle Q \rangle = \frac{\langle \psi_\theta  |Q|\psi_\theta  \rangle}{\langle \psi_\theta |\psi_\theta \rangle}
    \label{eq:appendix_vmc_expectation}
\end{equation}
where $|\psi_\theta \rangle$ is a variational many-body state defined by amplitudes $\psi_\theta(\mathbf{x})$ in a basis $\mathbf{x}$, while $Q$ is a generic operator like Hamiltonian or one-body density operator. The direct calculation of the expectation value in numerator and denominator cannot be performed exactly, however, estimates can be made through Monte Carlo integration. In general, we can rewrite the expectation in Eq.~\eqref{eq:appendix_vmc_expectation} as
\begin{equation}
    \langle Q \rangle = \frac{\int d\mathbf{x} \langle \psi_\theta | \mathbf{x}\rangle \langle \mathbf{x} | Q |  \psi_\theta \rangle}{\int d\mathbf{x'} |\psi_\theta(\mathbf{x})|^2} = \int d\mathbf{x} p_\theta(\mathbf{x})Q_L(\mathbf{x})
\end{equation}
where
\begin{equation}
    |\bar \varphi(\mathbf{x})|^2 = \frac{|\psi_\theta(\mathbf{x})|^2}{\int d\mathbf{x'} |\psi_\theta(\mathbf{x'})|^2}
\end{equation}
and the local observable function $Q_L(\mathbf{x})$ is defined as:
\begin{equation}
    Q_L(\mathbf{x}) = \frac{ \langle \mathbf{x} | Q |  \psi_\theta \rangle}{ \langle \mathbf{x} | \psi_\theta \rangle}
\end{equation}
The unbiased estimator to the expectation value of $Q_L$ can then obtained as the empirical average of the associated local observable $Q_L$ evaluated at configurations $\mathbf{x_i}$:
\begin{equation}
    \langle Q \rangle  = \EX_{\mathbf{x} \sim|\bar \varphi(x)|^2}\biggr[Q_L(\mathbf{x})\biggr]  \approx \frac{1}{N_s}\sum_{i=1}^{N_s} Q_L(\mathbf{x_i})  
\end{equation}
An analogous method for estimating observables in the context of QFTs was developed in Ref.~\cite{martynVariationalNeuralNetworkAnsatz2023}. The approach, referred to as VMC in Fock space, expresses the expectation value of an $n$-particle local observable function $Q^{\text{loc}}_n(\rR_n)$ for a quantum field state $| \Psi \rangle$, as
\begin{equation}
    \langle Q \rangle = \frac{\langle \Psi |Q|\Psi \rangle}{\langle \Psi|\Psi\rangle} =  \EX_{n \sim P_n} \EX_{\rR_n \sim |\bar \varphi_n|^2} \biggr[Q^{\text{loc}}_n(\rR_n) \biggr]\, ,
\end{equation}
where the n-particle local observable of the configuration $\rR_n = \{\rr_1, \rr_2, \dots, \rr_n\}$ is
\begin{equation}
    Q^{\text{loc}}_n(\rR_n) = \frac{ \langle \rR_n | Q |  \Psi \rangle}{ \langle \rR_n | \Psi\rangle} 
\end{equation}
with the probability distribution of the particle number
\begin{equation}
    P_n = \frac{ \langle \varphi_n | \varphi_n\rangle}{ \langle \Psi|\Psi \rangle} = \frac{\int \dd^n \rr |\varphi_n(\rR_n)|^2}{\sum_{m=0}^\infty \int \dd^n \rr' |\varphi_m(\rR'_m)|^2} 
    \label{eq:appendix_def_pn} 
\end{equation}
and the probability distribution of the n-particle wave function
\begin{equation}
    |\bar \varphi_n(\rR_n)|^2 = \frac{|\varphi_n(\rR_n)|^2}{\int \dd^n \rr' |\varphi_n(\rR'_n)|^2}\, .
    \label{eq:appendix_def_phi_bar}
\end{equation}
The above approach can be seen as a natural generalization of the traditional VMC, where the expectation value is extended to include both the distribution of the $n$-particle wave function and the particle number distribution. The unbiased estimator to the expectation value of $Q^{\text{loc}}_n(\rR_n)$ can then obtained as the empirical average of $Q^{\text{loc}}_n(\rR_n)$ evaluated at configurations $\rR_n$ drawn jointly from $P_n$ and $|\bar \varphi_n(\rR_n)|^2$:
\begin{equation}
    \langle Q \rangle  = \EX_{n \sim P_n} \EX_{\rR_n \sim |\bar \varphi_n|^2} \biggr[Q^{\text{loc}}_n(\rR_n) \biggr] \approx \frac{1}{\sum_n N_n}\sum_{n} \sum_{i_n=1}^{N_n} Q^{\text{loc}}_n(\rR_n^{(i_n)})
    \label{eq:appendix_exp_value}
\end{equation}
We next demonstrate that the developed VMC formalism in Fock space can be effectively applied to estimate key physical observables required for the optimization. Furthermore, in Appendix~\ref{appendix:estim_obsv}, we show that it is capable of estimating practically useful quantities, such as the one-body density matrix and the condensate fraction.

\section{Stochastic optimization}
The goal of VMC  is to minimize the energy expectation value with respect to the trial state $|\psi_\theta\rangle$ by finding the optimal variational parameters 
\begin{equation}
    \theta_{\text{opt}} = \argmin_{\theta} E(\theta) = \argmin_{\theta} \frac{\langle \psi_\theta|H|\psi_\theta \rangle}{\langle \psi_\theta|\psi_\theta \rangle} \geq E_0 \, ,
\end{equation}
where $E_0$ is the true ground state energy. Standard gradient-based optimization methods can be employed to determine the optimal parameters $ \theta_{\text{opt}} $ by iteratively updating the parameters using the gradient $ g = \nabla_\theta E(\theta) $, starting from an initial guess. However, a more general optimization approach, motivated by physical principles, is based on imaginary-time evolution. We next present an overview of this approach, which closely follows Ref.~\cite{medvidovicNeuralnetworkQuantumStates2024}, where the topic is discussed in greater detail.

Imaginary-time evolution is a well-known technique that exponentially suppresses the higher-energy components of an arbitrary trial state $ |\psi\rangle $, such that the ground-state component $ |0\rangle $ dominates in the long-time limit:
\begin{align}
        |\psi(\tau)\rangle = e^{-\tau H} |\psi\rangle &= \sum_n c_n  e^{-\tau E_n} |n\rangle \notag\\
        &\propto |0\rangle + \sum_{n > 0} \frac{c_n}{c_0}  e^{-\tau (E_n-E_0)} |n\rangle \xrightarrow{\tau \to \infty} |0\rangle
\end{align}
assuming that the trial state is not orthogonal to the ground state, such that $ c_0 = \langle 0 | \psi \rangle \neq 0 $.

We then consider the imaginary-time evolution of the trial state $ |\psi_\theta\rangle $ over a small time interval $ \delta \tau $, governed by the operator $ e^{-\delta \tau \hat{H}} $. This evolution is equivalent to an update of the variational parameters $ \theta \to \theta' = \theta + \delta \theta $, such that:
\begin{equation}
    |\psi_{\theta + \delta \theta} \rangle \approx e^{-\delta \tau H }|\psi_\theta \rangle
    \label{eq:imag_time_evol}
\end{equation}
Expanding both sides of Eq.~\eqref{eq:imag_time_evol} to first order in the small parameter increments and taking the limit $ \delta \tau \to 0 $, we obtain:
\begin{equation}
    S \dot \theta =  -g
    \label{eq:sr_equation}
\end{equation}
where $\dot \theta = d \theta / d\tau$ and
\begin{equation}
    S_{\mu \nu} = 2 \operatorname{Re} \big\{ \langle O_\mu^\dagger O_\nu \rangle - \langle O_\mu^\dagger \rangle \langle O_\nu \rangle \big\} = 2 \operatorname{Re} \big\{ \langle O_\mu^\dagger ( O_\nu - \langle O_\nu \rangle ) \rangle \big\}
    \label{eq:s_matrix}
\end{equation} 
\begin{equation}
    g_{\mu} = 2 \operatorname{Re} \big\{ \langle O_\mu^\dagger H \rangle - \langle O_\mu^\dagger \rangle \langle H \rangle \big\} = 2 \operatorname{Re} \big\{ \langle O_\mu^\dagger ( E - \langle E \rangle ) \rangle \big\}
    \label{eq:gradient}
\end{equation}
Here, $ g = \nabla_\theta E(\theta) $ denotes the energy gradient with respect to the variational parameters, while the matrix $ S $ is commonly referred to as the quantum geometric tensor (QGT) or the quantum Fisher information matrix~\cite{beccaQuantumMonteCarlo2017}. The indices $ \mu $ and $ \nu $ label the variational parameters and the expectation values $ \langle \cdot \rangle \equiv \langle \psi_\theta | \cdot | \psi_\theta \rangle / \langle \psi_\theta | \psi_\theta \rangle $ are evaluated with respect to the normalized trial state $ |\psi_\theta\rangle $ at imaginary time $ \tau $. Diagonal operators $O_\mu$ are defined as
\begin{equation}
    O_\mu |\psi_\theta \rangle = \partial_{\theta^\mu} |\psi_\theta \rangle \,.
\end{equation}
The gradient and the S-matrix defined in Eqs.~\ref{eq:s_matrix} and~\ref{eq:gradient} depend  only on the local energy $ E^{\text{loc}}(\rR_n) $ and the local operator $ O^{\text{loc}}_\nu(\rR_n) $, which are sampled using the MCMC technique described in the previous section. The corresponding expectation values $ \langle E \rangle $ and $ \langle O_\nu \rangle $ can be further evaluated as empirical means over all sampled configurations $\rR_n$. This enables the estimation of the gradient $ g $ and the S-matrix required for optimizing the model parameters, without the need for direct evaluation of $ \nabla_\theta E(\theta) $.

Since the  Eq.~\eqref{eq:sr_equation} reduces to the explicit inversion of the S-matrix (known as \textit{stochastic reconfiguration} (SR) update rule), the approach is computationally expensive for large models, as it requires solving an $ N_p \times N_p $ linear system at each iteration, where $ N_p $ denotes the number of model parameters. To mitigate this cost, one can employ a computational trick introduced in Ref.~\cite{chenEmpoweringDeepNeural2024}. Specifically, by utilizing the Moore-Penrose pseudoinverse, Eq.~\eqref{eq:sr_equation} can be reformulated as
\begin{equation}
    \dot \theta = S^{-1}g = (\bar O^\dag \bar O)^{-1} \bar O^\dag \bar \epsilon =  \bar O^\dag (\bar O\bar O^\dag )^{-1} \bar \epsilon \; .
    \label{eq:min_sr}
\end{equation}
In the final expression, the rank of the S-matrix is at most $ N_s $ and, consequently, the matrix size reduces from $ N_p \times N_p $ to $ N_s \times N_s $. This method is therefore particularly useful for large models where the number of parameters $ N_p $ significantly exceeds the total number of samples per iteration $ N_s $. Following the Ref.~\cite{martynVariationalNeuralNetworkAnsatz2023}, we introduce \textit{the neural tangent kernel}, defined as:
\begin{equation}
    T = \bar{O} \bar{O}^\dagger
\end{equation}
where we defined centered and normalized variables:
\begin{equation}
    \bar{\epsilon} =  (E^{\text{loc}}(\rR_n) - \langle E \rangle)/\sqrt{N_s}
    \label{eq:eps_bar}
\end{equation} 
\begin{equation}
   \bar{O}_\nu =   (O^{\text{loc}}_\nu(\rR_n) - \langle O_\nu \rangle)/\sqrt{N_s} 
   \label{eq:o_bar}
\end{equation}
while the term $O^{\text{loc}}_\nu(\rR_n)$ can be directly evaluated as

\begin{align}
        O^{\text{loc}}_\nu(\rR_n) 
        &= \frac{1}{ \langle \rR_n | \Psi\rangle} \biggr \langle \rR_n  \biggr |\int \dd \rr \hat \psi^\dag(\rr ) \partial_\nu \hat \psi(\rr )  \biggr|\Psi \biggr \rangle \notag \\
    &=  \frac{1}{\phi_n(\rR_n)}\partial_\nu \phi_n(\rR_n) =  \partial_\nu \ln \phi_n(\rR_n).
\end{align}

For the Hamiltonian under consideration, the local energy $E^{\text{loc}}(\rR_n)$ can be decomposed into distinct contributions, including the kinetic energy, potential energy, and interaction energy terms:
\begin{equation}
    E^{\text{loc}}(\rR_n) =  \frac{ \langle \rR_n | H |  \Psi \rangle}{ \langle \rR_n | \Psi\rangle} = E^{\text{loc}}_{\text{kin}}(\rR_n)  + V^{\text{loc}}_{\text{ext}}(\rR_n)   + W^{\text{loc}}_{\text{2-body}}(\rR_n)  
\end{equation}
The expectation value of the local kinetic energy may be expressed as
\begin{align}
    E^{\text{loc}}_{\text{kin}}(\rR_n)  &= \frac{1}{ \langle \rR_n | \Psi\rangle} \biggr \langle \rR_n  \biggr |\int \dd\rr \hat \psi^\dag(\rr ) \biggr(\frac{-1}{2m}\nabla^2 \biggr) \hat \psi(\rr )  \biggr|\Psi \biggr \rangle \notag \\
    &=  \left( -\frac{1}{2m} \right) \frac{1}{\phi_n(\rR_n)}\nabla^2 \phi_n(\rR_n) \notag \\
    &=  -\frac{1}{2m} \left( \nabla^2 \ln \phi_n(\rR_n) + \left( \nabla \ln \phi_n(\rR_n) \right)^2 \right)
\end{align}
The expectation value of the local external and chemical potential terms may be expressed as
\begin{align}
    V^{\text{loc}}_{\text{ext}}(\rR_n)  &= \frac{1}{ \langle \rR_n | \Psi\rangle} \biggr \langle \rR_n  \biggr |\int d\rr \hat \psi^\dag(\rr)\biggr(V(\rr) - \mu \biggr) \hat \psi (\rr) \biggr|\Psi \biggr \rangle  \notag \\
     &= \frac{1}{ \phi_n(\rR_n)}\sum_{i=1}^n \biggr(V(\rr_i) - \mu \biggr)  \phi_n(\rR_n) = \sum_{i=1}^n \biggr(V(\rr_i) - \mu \biggr)\, ,
\end{align}
while the expectation value of the local two-body interaction potential term is
\begin{align}
        W^{\text{loc}}_{\text{2-body}}(\rR_n)   &=   \frac{1}{ \langle \rR_n | \Psi\rangle} \biggr \langle \rR_n  \biggr |\int \dd\rr \dd\rr' \hat \psi^\dag(\rr)\hat \psi^\dag(\rr') W(\rr - \rr') \hat \psi (\rr') \hat \psi(\rr)\biggr|\Psi \biggr \rangle \notag \\
         &= \frac{1}{ \phi_n(\rR_n)}\sum_{i<j}^n W(\rr_i - \rr_j)  \phi_n(\rR_n) = \sum_{i<j}^n W(\rr_i - \rr_j)
\end{align}

The derived expression requires computing the gradient and Laplacian of the logarithm of the n-particle wave function amplitude, which is provided as output of our ansatz.

If we approximate $ S \approx \mathbb{I} $ in Eq.~\eqref{eq:sr_equation}, we obtain the standard \textit{stochastic gradient descent} (SGD) optimization rule:
\begin{equation}
    \theta'= \theta - \delta \tau \nabla_\theta  E(\theta)
    \label{eq:sgd}
\end{equation}
Now, we can use any of the popular optimizers from deep learning, such as Adam~\cite{kingmaAdamMethodStochastic2015}. This method is the least computationally demanding and was employed in Ref.~\cite{martynVariationalNeuralNetworkAnsatz2023}. However, it often converges to higher ground state energies compared to the SR (minSR) methods, or may even fail to converge for large systems.

\section{MCMC in Fock space}
\label{appendix:mcmc_fock}
To generate the samples $\{\rR_i: i = 1, \dots, N_s\}$ required for evaluating the observables in Eq.~\eqref{eq:appendix_exp_value}, it is standard practice to employ established Markov Chain Monte Carlo (MCMC) methods. In the case of Fock-space VMC, the algorithm must be generalized to sample jointly from $ P_n $ and $ |\bar{\varphi}_n|^2 $. This extension can be regarded as a generalization of standard MCMC sampling in the classical grand-canonical ensemble to the quantum grand-canonical ensemble. In this work, we largely follow the sampling strategy introduced in Ref.~\cite{martynVariationalNeuralNetworkAnsatz2023}, extending it to the case of $d \geq 1$. 

As in conventional MCMC algorithms, the procedure in Fock space begins by proposing a new configuration $\rR_n'$ from the current configuration $\rR_n$. The proposed configuration is then accepted or rejected according to the Metropolis–Hastings probability:
\begin{equation}
    p_{\text{accept}} = \min\Biggl(1, \frac{P_\text{bos}(\rR_n')}{P_\text{bos}(\rR_n)} \frac{g(\rR_n | \rR_n')}{g(\rR_n' | \rR_n)} \Biggr),
\end{equation}
where $g(\rR_n' | \rR_n)$ denotes the transition probability from $\rR_n \rightarrow \rR_n'$ under the chosen proposal scheme, and the probability of a configuration $\rR_n$ is given by
\begin{align}
    P(\rR_n) &= P_n \cdot |\bar{\varphi}_n|^2 \dd^n \rR\notag \\
    &=  \frac{\int \dd^n \rr' |\varphi_n(\rR'_n)|^2}{\sum_{m} \int \dd^m \rr' |\varphi_m(\rR'_m)|^2}  \cdot \frac{|\varphi_n(\rR_n)|^2}{\int \dd^n \rr' |\varphi_n(\rR'_n)|^2} \dd^n \rR\notag \\
    &= \frac{|\varphi_n(\rR_n)|^2 \dd^n\rR}{\sum_m \int \dd^m \rr' |\varphi_m(\rR'_m)|^2},
\end{align}
where the definitions of $ P_n $ from Eq.~\eqref{eq:appendix_def_pn} and $ \bar{\varphi}_n $ from Eq.~\eqref{eq:appendix_def_phi_bar} have been used.

Because we focus on bosonic states, each configuration $\rR_n$ is physically indistinguishable from any of its permutations $\rR_n'' \in \text{Sym}(\rR_n)$. Accordingly, the probability associated with a bosonic configuration $\rR_n$ is
\begin{equation}
    P_\text{bos}(\rR_n) = \sum_{\rR''_n \in \text{Sym}(\rR_n)}  P(\rR''_n) = \frac{n!|\varphi_n(\rR_n)|^2 \dd^n \rr }{\sum_m \int \dd^m \rr' |\varphi_m(\rR'_m)|^2}.
\end{equation}

We next specify the proposal scheme, which enables us to generate new bosonic configurations $ \rR_n' $ and also allows for changes in particle number. With probability $2p_\pm$, the scheme proposes configurations with particle number increased or decreased, each with equal probability $p_\pm$. With the remaining probability $p_0 = 1 - 2p_\pm$, the particle number is preserved, and a uniform random displacement is applied to all particle positions:
\begin{equation}
    \rR_n \rightarrow \rR_n' = \rR_n + \bm{\zeta},
\end{equation}
where $\bm{\zeta} \sim U\left([ -\tfrac{w}{2}, \tfrac{w}{2} ]^d\right)$ is a $d$-dimensional uniform random vector of width $w$. In this case only, the transition rule is symmetric, i.e., $ g(\rR_n | \rR_n') = g(\rR_n' | \rR_n) $, so the Metropolis–Hastings acceptance probability for the transition $\rR_n \rightarrow \rR_n'$ reduces to
\begin{equation}
    p_{\text{accept}} = \min\Biggl(1, \frac{|\varphi_n(\rR_n')|^2}{|\varphi_n(\rR_n)|^2} \Biggr).
\end{equation}

On the other hand, with probability $p_\pm$, the proposal function either adds or removes a particle. In the case of particle addition, a new particle is introduced at a position $\rr_{n+1}$, which is sampled uniformly at random from the domain $[0, L]^d$. With the same probability $p_\pm$, a particle is removed by selecting an index $i \in \{1, \dots, n\}$ uniformly at random. These proposals are not symmetric: for particle addition, the proposal probability is 
\begin{equation}
g(\rR_{n+1} | \rR_n) = p_\pm \cdot \frac{\dd\rr_{n+1}}{L^d},
\end{equation}
while for particle removal it is 
\begin{equation}
g(\rR_{n-1} | \rR_n) = p_\pm \cdot \frac{1}{n}.
\end{equation}
Hence, the Metropolis–Hastings acceptance probability for adding a particle, i.e., for the transition $\rR_n \rightarrow \rR_{n+1}$, is
\begin{align}
p_{\text{accept}} &= \min\left(1, \frac{P_{\text{bos}}(\rR_{n+1})}{P_{\text{bos}}(\rR_n)} \cdot \frac{g(\rR_n | \rR_{n+1})}{g(\rR_{n+1} | \rR_n)}\right) \notag \\
&= \min\left(1, \frac{(n+1)! \, |\varphi_{n+1}(\rR_{n+1})|^2 \, \dd^{n+1}\rr}{n! \, |\varphi_n(\rR_n)|^2 \, \dd^n \rr} \cdot \frac{p_\pm}{n+1} \cdot \frac{L^d}{p_\pm \, \dd\rr_{n+1}}\right) \notag \\
&= \min\left(1, \frac{L^d \cdot |\varphi_{n+1}(\rR_{n+1})|^2}{|\varphi_n(\rR_n)|^2}\right) \, ,
\end{align}
while the acceptance probability for removing a particle, i.e., $\rR_n \rightarrow \rR_{n-1}$, is
\begin{align}
p_{\text{accept}} &= \min\left(1, \frac{P_{\text{bos}}(\rR_{n-1})}{P_{\text{bos}}(\rR_n)} \cdot \frac{g(\rR_n | \rR_{n-1})}{g(\rR_{n-1} | \rR_n)}\right) \notag \\
&= \min\left(1, \frac{(n-1)! \, |\varphi_{n-1}(\rR_{n-1})|^2 \, \dd^{n-1}\rr}{n! \, |\varphi_n(\rR_n)|^2 \, \dd^n \rr} \cdot \frac{p_\pm \, \dd\rr_{n-1}}{L^d} \cdot \frac{n}{p_\pm}\right) \notag \\
&= \min\left(1, \frac{1}{L^d} \cdot \frac{|\varphi_{n-1}(\rR_{n-1})|^2}{|\varphi_n(\rR_n)|^2}\right) \, .
\end{align}

We typically set $p_\pm = 0.25$, so that configurations with different particle numbers are proposed with probability $2p_\pm = 50\%$. A larger value of $ p_\pm $ will force the algorithm to explore more particle-number sectors, which could significantly decrease the acceptance ratio. On the other hand, a smaller value of $p_\pm$ enhances the algorithm's ability to find optimal configurations within the same particle-number sector.

In practice, we run around 100 MCMC chains in parallel, using a sweep size (i.e., the number of Metropolis–Hastings steps performed before accepting a configuration as a valid sample) of 30 and collect 50 to 100 samples per chain. Expectation values are then estimated as empirical means over all collected samples. 

\section{Estimation of observables}
\label{appendix:estim_obsv}
To verify that the ansatz correctly captures the ground state of the system, we compare not only the ground-state energy and particle number but also employ VMC to estimate additional observables, such as the ground-state particle-number density $n(\rr)$ and the one-body density matrix $\rho(\rr, \rr')$. In this section, we derive expressions for the observables analyzed in the main text.

\subsection{Particle Number Density}
The particle number density can be expressed as:
\begin{align}
    n(\rr) &= \frac{1}{\langle \Psi | \Psi \rangle} \langle \Psi |\hat \psi^\dag (\rr) \hat \psi (\rr)|\Psi \rangle 
    = \frac{1}{\langle \Psi | \Psi \rangle} \langle \Psi | \int \dd\rr'\; \delta(\rr' - \rr)\hat \psi^\dag (\rr') \hat \psi (\rr')|\Psi \rangle \notag \\
    &= \frac{1}{\langle \Psi | \Psi \rangle} \sum_{n=0}^\infty \int \dd^n\rr\; |\varphi_n(\rr_n)|^2\biggr(\sum_{i=1}^n\delta(\rr - \rr_i)\biggr) = \frac{1}{\langle \Psi | \Psi \rangle} \sum_{n=0}^\infty \sum_{i=1}^n \int \dd^{n \backslash i}\rr\; |\varphi_n(\rr_n|\rr_i = \rr)|^2 \notag \\
   &= \EX_{n \sim P_n} \left( \sum_{i=1}^n  \EX_{\rr_{n  \backslash i} \sim |\bar \varphi_{n  \backslash i}|^2} \left[\frac{|\varphi_n(\rr_n|\rr_i = \rr)|^2}{\int d\rr_i |\varphi_n(\rr_n)|^2} \right] \right),
\end{align}
where the second expectation value is taken over the marginal distribution $|\bar \varphi_{n  \backslash i}|^2 = \int \dd\rr_i |\bar \varphi_n|^2$, and the notation $\varphi_n(\rR_n|\rr_i = \rr)$ means that this function is evaluated at $\rR_n$ with the constraint that $\rr_i = \rr$. Samples from the marginal distribution may be obtained by sampling $\rR_n \sim  |\bar \varphi_n|^2$ as usual, and ignoring $\rr_i$.

\subsection{One-body Density Matrix}
In studying the ground state of a bosonic system, another very useful observable is the so-called one-body density matrix (OBDM), $\rho(\rr, \rr')$, which can be expressed in terms of the particle creation and annihilation operators as
\begin{equation}
    \rho(\rr, \rr') = \frac{1}{\langle \Psi | \Psi \rangle} \langle \Psi |\hat \psi^\dag (\rr') \hat \psi (\rr)|\Psi \rangle
\end{equation}
At zero-temperature, off-diagonal long-range order (ODLRO) is a key indicator of the Bose-Einstein condensation (BEC) phase. In the presence of ODLRO, the OBDM of the ground state $|\Psi_0 \rangle$, approaches a nonzero plateau for large separations, that is, $\lim_{|\rr - \rr'| \rightarrow \infty} \rho(\rr, \rr') = \rho_0$ where $\rho_0$ is the condensate density~\cite{penroseBoseEinsteinCondensationLiquid1956a}.

\subsubsection{Translation-invariant system}
In a translationally invariant system, the OBDM depends only on the relative displacement vector $\bm{s} = |\rr - \rr'|$:
\begin{align}
        \rho(\bm{s}) &=  \frac{1}{\langle \Psi | \Psi \rangle} \sum_{n=0}^{\infty} n \int \dd^{n-1}\rr \varphi_n^*(\rr_1 + \bm{s}, \rr_2, ..., \rr_N)\varphi_n(\rr_1, \rr_2, ..., \rr_n)  \notag \\ 
        &=  \frac{1}{\langle \Psi | \Psi \rangle} \sum_{n=0}^{\infty} \frac{n}{L^d} \int \dd^{n}\rr \varphi_n^*(\rr_1 + \bm{s}, \rr_2, ..., \rr_N)\varphi_n(\rr_1, \rr_2, ..., \rr_n)  \notag \\ 
        &= \EX_{n \sim P_n} \EX_{\rr_n \sim |\bar \varphi_n|^2} \biggr[\frac{n}{L^d} \frac{\varphi_n^*(\rr_n | \rr_1 = \rr_1 + \bm{s})}{\varphi_n(\rr_n)} \biggr].
\end{align}
This expression can now be straightforwardly evaluated as an empirical mean over configurations $\rr_n$ sampled jointly from $P_n$ and $|\bar \varphi_n|^2$.

\subsubsection{Inhomogeneous system}

In the absence of translational invariance, the analysis becomes considerably more complex. The OBDM, expressed as an expectation value, is given by:
\begin{align}
        \rho(\rr, \rr') &= \frac{1}{\langle \Psi | \Psi \rangle} \langle \Psi |\hat \psi^\dag (\rr) \hat \psi (\rr')|\Psi \rangle \notag \\
        &=  \frac{1}{\langle \Psi | \Psi \rangle} \sum_{n=0}^{\infty} n \int \dd^{n-1}\rr \varphi_n^*(\rr, \rr_2, ..., \rr_N)\varphi_n(\rr', \rr_2, ..., \rr_n)  \notag \\ 
        &= \EX_{n \sim P_n} \EX_{\rr_n \sim |\bar \varphi_n|^2} \biggr[n \frac{\varphi_n^*(\rr_n| \rr_1 = \rr)\varphi_n(\rr_n| \rr_1 = \rr')}{\int   d\rr_1 |\varphi_n(\rr_n)|^2} \biggr] 
\end{align}

A direct evaluation of the above expression involves numerical integration over a $2 \cdot d$-dimensional grid, which becomes computationally challenging for $d = 2$. One possible approach to avoid the direct integration is to employ the so-called \textit{natural orbitals}, which play a role similar to marginal distributions for probability densities~\cite{pfauAccurateComputationQuantum2024}. The idea is based on projecting the OBDM into an orthogonal basis set $\phi_1(\rr), ..., \phi_M(\rr)$, such that elements of the projected OBDM, $\rho_{ij}$, are given by
\begin{equation}
    \rho_{ij} = \int d\rr d\rr'\phi_i(\rr)\phi_j(\rr')\rho(\rr, \rr')
\end{equation}
In the case of a single state, a Monte Carlo estimator for $\rho_{ij}$ can be written as~\cite{pfauAccurateComputationQuantum2024}:
\begin{equation}
     \langle \hat \rho_{ij} \rangle = \EX_{\rr_n\sim |\bar \varphi_n|^2} \EX_{\rr' \sim \rho}\biggr[\frac{\phi_i(\rr)\phi_j(\rr')\varphi_n( \rr', \rr_2, ..., \rr_n)}{\rho(\rr') \varphi_n( \rr, \rr_2, ..., \rr_n)}\biggr]
     \label{eq:obdm_inhom_sampling}
\end{equation}
where $\rho$ is an arbitrary probability distribution over single particle states. While any $\rho$ can work in principle, selecting a distribution that more closely approximates the actual boson density results in an estimator with reduced variance. For instance, in the case of bosons confined in a two-dimensional harmonic potential (see Sec.~\ref{subsec:harm_confined}), an appropriate choice for the orthonormal basis would be the set of eigenfunctions of the corresponding single-particle Hamiltonian and a practical choice for $\rho$ is the ground-state density of a single harmonically trapped boson:
\begin{equation}
    \rho(\rr) = \frac{\beta^2}{\pi}e^{-\beta^2\rr^2}.
\end{equation}

\section{Convergence Analysis}
\label{appendix:rescaled_var}

The primary observable tracked during the VMC procedure is the total energy that we estimate as empirical means over all collected samples:
\begin{equation}
    \bar{E}  = \EX_{n \sim P_n} \EX_{\rR_n \sim |\bar \varphi_n|^2} \left[E^{\text{loc}}_n(\rR_n) \right] \approx \frac{1}{N_s} \sum_{i =1}^{N_s} E^{\text{loc}}_n(\rR_n^{(i)}) 
\end{equation}
The uncertainty in the estimated expectation value is computed as the standard deviation of the means across the chains (standard error):
\begin{equation}
    \Delta E = \frac{1}{N_c} \sqrt{\sum_{c=1}^{N_c} \left(\bar{E}_c - \bar{E}\right)^2}
\end{equation}
where $N_c$ is the number of MCMC chains, and $\bar{E}_c$ is the mean local energy of the $c$-th chain. 

\begin{figure}[!t]
    \centering
    \includegraphics[width=.7\linewidth]{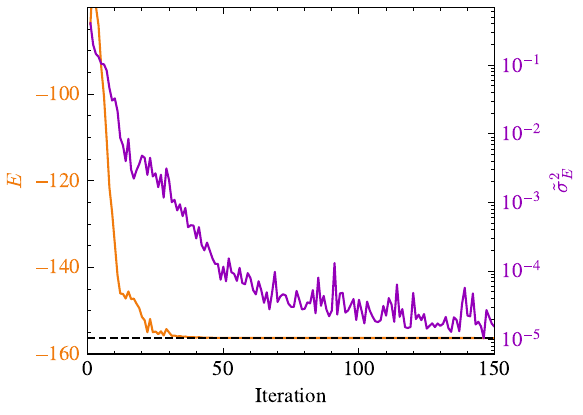}
    \caption{Energy (left axis) and rescaled variance (right axis) vs.\ iteration during the optimization of the TF ansatz applied to the one-dimensional Calogero-Sutherland model with the parameters studied in the main text ($\mu = 3 \cdot 5^2 \cdot \frac{\pi^2 \lambda^2}{6 m L^2}$, $g = 5.0$). The exact ground-state energy is shown as a black dashed line.}
    \label{fig:TF_LL_1D_variance}
\end{figure}

Additionally, to analyze convergence during optimization and evaluate the resulting performance of the architectures, we introduce the \textit{rescaled variance} parameter, defined as:
\begin{equation}
    \tilde{\sigma}^2_E = N \frac{\langle H^2 \rangle - \langle H\rangle^2}{\langle H\rangle^2}
\end{equation}
This metric is closely related to the V-score, which is defined for spin systems and is known to correlate with the accuracy of the wave function~\cite{wuVariationalBenchmarksQuantum2024}. Like the V-score, the rescaled variance provides a reliable measure of wave function accuracy and can be computed entirely independently of exact reference energies. Figure~\ref{fig:TF_LL_1D_variance} shows the energy along with the rescaled variance $\tilde{\sigma}^2_E$ during the optimization of the TF ansatz applied to the one-dimensional Calogero-Sutherland model. The plot illustrates how the energy converges to the exact ground-state value, while the rescaled variance simultaneously drops to a value on the order of $10^{-5}$, confirming a good approximation of the true ground state.

\section{Two-dimensional Calogero–Sutherland model}
\label{appendix:cs_2d}

The exact ground state of the Calogero–Sutherland model in arbitrary dimensions was studied in Ref.~\cite{khareQuantumManybodyProblem1997}. Here, we present an overview of the specific two-dimensional case considered in the main text. The ground-state eigenfunction and eigenvalue of the two-dimensional Hamiltonian are given by:
\begin{equation}
    \Psi_0 = \hat{C} \exp\left(-\frac{\omega}{2} \sum_{i=1}^{N} r_i^2\right) \prod_{i<j} |\vb{r}_i - \vb{r}_j|^{\Lambda_2},
    \label{eq:psi0_cs2d}
\end{equation}
\begin{equation}
    E_0 = [2N + N(N - 1)\Lambda_2]\omega,
\end{equation}
where $\omega$ denotes the frequency of the harmonic trap, and we have chosen units such that $\hbar = 2m = 1$. The parameter $\Lambda_2$ is defined as:
\begin{equation}
    \Lambda_2 =  \sqrt{G/2} =  \sqrt{g/2}.
\end{equation}

By expressing the position vectors $\vb{r}_i \in \mathbb{R}^2$ as complex numbers $z_i \in \mathbb{C}$, the ground-state probability density can be rewritten as:
\begin{equation}
    |\Psi_0|^2 = C \exp\left(- \Lambda_2 \sum_{i=1}^{N} |z_i|^2\right) \prod_{i<j} |z_i - z_j|^{2\Lambda_2}.
    \label{eq:psi0_square_cs2d}
\end{equation}

For the special case of $\Lambda_2 = 1$ (which corresponds to $g = G = 2$), the density $|\Psi_0|^2$ in Eq.~\eqref{eq:psi0_square_cs2d} takes the exact form of the joint probability density function for the eigenvalues of matrices drawn from an ensemble of complex random matrices~\cite{mehta1991random}. Following Ref.~\cite{mehta1991random}, one can systematically derive the $n$-point correlation functions for all $n$. In particular, the one-point correlation function, interpreted as the particle number density, is given by:
\begin{equation}
    \mathcal{R}(r) = \frac{\omega}{\pi \Lambda_2} \exp(-\omega r^2) \sum_{p=0}^{N-1} \frac{(\omega r^2)^p}{p!}.
\end{equation}
As shown in Fig.~\ref{fig:TF_CS_2D} of the main text, this density profile is completely isotropic and does not depend on the angular coordinate.

\end{document}